\documentclass[fleqn,usenatbib]{mnras}
\makeatletter
\@ifundefined{@parse@version@dash}{%
\def\@parse@version#1{\@parse@version@0#1}
\def\@parse@version@#1/#2/#3#4#5\@nil{%
\@parse@version@dash#1-#2-#3#4\@nil}
\def\@parse@version@dash#1-#2-#3#4#5\@nil{%
  \if\relax#2\relax\else#1\fi#2#3#4 }
}{}
\makeatother

\usepackage[T1]{fontenc}

\DeclareRobustCommand{\VAN}[3]{#2}
\let\VANthebibliography\thebibliography
\def\thebibliography{\DeclareRobustCommand{\VAN}[3]{##3}\VANthebibliography}

\usepackage{graphicx}	
\usepackage{amsmath, xcolor}	
\usepackage{amssymb, tabularx}	

\title[IRAS 18144--1723 maser observations]{Characterising the high-mass star forming region IRAS 18144--1723 through methanol maser observations}

\author[E. Khafagy et al.]{
\Large{Khafagy Esraa$^{1}$, \thanks{E-mail: esraakhafagy@gstd.sci.cu.edu.eg}
Edris K.A.$^{2}$,
Shalabiea O.M.$^{1,3}$,
Bartkiewicz A.$^4$, Richards A.M.S.$^5$ 
and Awad Z.$^{1}$} \\
$^1$Astronomy, Space Science and Meteorology Department, Faculty of Science, Cairo University, Giza 12613, Egypt.\\
$^2$Astronomy and Meteorology Department, Faculty of Science, Al-Azhar University, Cairo, Egypt.\\
$^3$Faculty of Navigation Science and Space Technology, Beni-Suef University, Beni-Suef, Egypt.\\
$^4$Institute of Astronomy, Faculty of Physics, Astronomy and Informatics, Nicolaus Copernicus University, Grudziadzka 5, 87-100 Torun, Poland.\\
$^{5}$Jodrell Bank Centre for Astrophysics, Department of Physics $\&$ Astronomy, The University of Manchester, M13 9PL, UK.\\
}

\date{Accepted XXX. Received YYY; in original form ZZZ}

\pubyear{2022}

\begin{document}
\label{firstpage}
\pagerange{\pageref{firstpage}--\pageref{lastpage}}
\maketitle

\begin{abstract}

We introduce a study of the massive star forming region IRAS 18144--1723 using  observations of the 6.7 GHz methanol maser line. Such regions are opaque at short wavelengths but can be observed through radio emission lines. In this study  we traced the kinematics of the source on milliarcsecond scales using the Multi-Element-Radio-Interferometer-Network (MERLIN). We found 52 maser spots in the LSR velocity range 45--52 km s$^{-1}$, near the centre of the previously-detected CO range of 21.3--71.3 km s$^{-1}$, lying within $\sim$ 0$''$.5 of IRAS 18144--1723 `B', thought to be a young Class I protostar.
Their distribution can be approximated as an ellipse, which, if it were rotating, would have its axis oriented south-east to north-west.  The most probable morphology of the emitting regions is interaction between a disc and an outflow, possibly with a very large opening angle. The arcmin-scale CO outflow centred on source `B' is oriented East-West, and the methanol masers do show the highest dispersion of velocity gradients in approximately this direction, so the kinematics are complex and suggest that more than one source may be responsible. We also tested kinematic models for a Keplerian disc or a simple bipolar outflow, but neither are compatible with the kinematics of the maser clumps and the characteristics of their internal velocities.

\end{abstract}

\section{Introduction}
The evolution of high mass stars {is} of extreme importance in astrophysics, due to its influence on galaxy formation and the interstellar medium,  as high-mass stars are the main source of heavy elements and other materials feeding into the next generation of new born stars.
There are some difficulties faced during the study of high-mass star formation; most are at large distances ($>$1 kpc); they are formed in high density clouds  and radiation pressure from massive proto-stellar objects tends to drive material away \citep{Bonnell1998}.
 There are many theories on how massive stars form; however the most acceptable two theories, so far, are those in which the first suggests an extension of the core-accretion mechanism for low-mass star formation \citep{McKee2003} and the second proposes a possible merging of low and intermediate mass stars \citep{Keto2010}.

The latter theory is supported by observation of massive starless cores which have mass functions similar to the stellar initial mass function (IMF) for low-mass stars (e.g. \citealt{Beuther2004}; \citealt{Reid2006}). The extended core-accretion theory explained the differences between massive and less-massive stars including that massive stars have more energetic outflows, evolve faster and that the hydrogen core starts burning before accretion finishes \citep{Arce2007}. Recent evidence for massive star formation supports core accretion, through two scenarios: firstly monolithic core accretion and secondly competitive accretion starting from smaller condensations. In the second scenario, cores continue to accrete material from the surrounding cloud, probably via filaments, where the infall is orthogonal to any protostellar outflow \citep{Kong2019, Kong2021}.
These  studies suggest that the accretion disc angular momentum is influenced by the filaments as well as the (growing) enclosed mass, which might lead to more complex disc characteristics than for a disc around a core formed promptly from a single collapse.

\begin{table*}
\centering
\caption{Parameters of sources observed.}
\label{tab:obs}
\begin{tabular}{lllccccc}
\hline
Source & Role & \multicolumn{2}{c}{Position (J2000)} & Flux density & Frequency &$\!\!$Bandwidth$\!\!$ &$\!\!$ Channel width$\!\!$\\
       &      & $\alpha$ & $\delta$  & (Jy)  &(MHz) &  (MHz) &(kHz)\\
\hline
18144--1723 & target    & 18$^{\text{h}}$17$^\text{m}$24$^\text{s}\!$.4000 & --17$^{\circ}$22$'$13\farcs000 & -- & 6668.519$^{\dag}$ &1&3.90625\\
4C39.25    & bandpass, phase offset & 09$^{\text{h}}$27$^\text{m}$03$^\text{s}\!$.0139 & \,\,39$^{\circ}$22$'$20\farcs852 & 11.250 &6668.200 & 1 & 3.90625\\
1757--150   & phase calibrator&  17$^{\text{h}}$59$^\text{m}$57$^\text{s}\!$.2466 & --15$^{\circ}$52$'$49\farcs171&0.108&6668.400 &16 &500\\
4C39.25    & bandpass, phase offset& 09$^{\text{h}}$27$^\text{m}$03$^\text{s}\!$.0139 & \,\,39$^{\circ}$22$'$20\farcs852&11.250 &6668.400&16&500\\
\hline
\end{tabular}
\flushleft
$^\dag$ Rest frequency; the observing frequency was adjusted in the correlator to correspond to $V_{\mathrm{LSR}}$ = 56 km s$^{-1}$.\\
\end{table*}

High mass star forming regions are extremely opaque at visible wavelengths and hard to detect at near infrared (NIR) or even longer wavelengths. Masers are considered one of the best tracers for massive star forming regions \citep{Norris1998}. They characterise discs, bipolar outflows, magnetic fields and internal motion \citep{Gray2012}. The 6.7 GHz methanol masers occur during the early stages of massive star formation  but after the onset of outflows, implying accretion \citep{deVilliers2015}. The associated radiation and possibly shocks evaporate methanol from icy grain mantles formed during the initial stages of cloud collapse. We trace 6.7 GHz methanol masers towards the main source IRAS 18144--1723, an infrared source in a massive star forming region. The equatorial position of the source is ($\alpha$ = 18$^{\text{h}}$17$^\text{m}$24$^\text{s}\!$.4, $\delta$ = --17$^{\circ}$22$'$13$''$) \citep{IRAS1988}, within an uncertainty ellipse of ($28''$, $5''$) at a position angle 88$^{\circ}$. All  positions given in this work are in the J2000 equatorial coordinate system, given as {$\alpha$ in (h : m : s) and $\delta$ in ($^{\circ}~:~ ' ~: ~''$)}.

\citet{Molinari1996} {observed IRAS 18144--1723 and} detected an ammonia core giving a systemic velocity of 47.3 km s$^{-1}$ and used this to deduce a kinematic distance of 4.33 kpc. We tested this distance with the Bessel project on-line calculator\footnote{http://bessel.vlbi-astrometry.org}, which is based on accurate distances and proper motions of high mass star forming regions across the Milky Way \citep{Reid2019}, {and obtained a distance of 3.9 $\pm$ 0.22 kpc which we used in our calculations throughout this study. This new distance is in good agreement with the distance range originally obtained by \citet{Molinari1996} within errors}.
The 6-cm emission which was detected 93 arcsec away from the IRAS source location is not likely to be associated \citep{Molinari1998}. The upper limit towards IRAS 18144--1723, (3$\sigma_{\mathrm{rms}}~\sim$ 0.2 mJy at wavelength 6 cm), implies an electron density and emission measure two orders of magnitude lower than the lower limit typical for UCH{\sc ii} regions \citep{Kurtz2005}, suggesting that IRAS 18144--1723 is in a pre-UCH{\sc ii} stage. \citet{Zhang2005} did not detect CO J=2--1 from this region. \citet{Varricatt2010} detected an IR K-band (2.2 $\mu$m) source `A' within $1\farcs3$ of the IRAS position. They also detected H$_2$ lines. The continuum-subtracted H$_2$ image showed a bow-shock at (18$^{\text{h}}$17$^{\text{m}}$23$^{\text{s}}\!$.13, --17$^{\circ}$22$'$13\farcs5) almost due West of `A' (position angle 274$^{\circ}$) indicating an East--West (E--W) outflow with a collimation factor of nearly 10. \citet{Varricatt2018}, {\it hereafter} V18, succeeded in detecting CO J = 3--2 in three isotopologues ($^{12}$CO, $^{13}$CO and C$^{18}$O), showing a bipolar outflow extending in an E--W direction either side of the IRAS source. V18 observed the IRAS source using UIST IR imaging at K, L$'$ and M$'$ bands  at 2.2, 3.77 and 4.69 microns, respectively, and longer-wavelength filters. They found that the source can be resolved into two components ; `A' (18$^{\text{h}}$17$^{\text{m}}$24$^{\text{s}}\!$.375, --17$^{\circ}$22$'$14\farcs71) and `B' (18$^{\text{h}}$17$^{\text{m}}$24$^{\text{s}}\!$.239, --17$^{\circ}$22$'$12\farcs87), respectively, with an astrometric accuracy $\sim 0\farcs5$, {which is equivalent to a projected separation of 10140 au at 3.9 kpc}. Source `A' was  not detected in the short wavelength bands J and H, and source `B'  was  seen clearly using the M band and longer wavelengths filters, suggesting that `B' is even more obscured than `A'. V18 detected slightly resolved emission from the region using SCUBA-2 at 450 $\mu$m and 850 $\mu$m with full width half maxima of 18\farcs5 and 22\farcs6, respectively, peaking at {(18$^{\text{h}}$17$^{\text{m}}$24$^{\text{s}}\!$.010, --17$^{\circ}$22$'$12\farcs05), closer to `B' than to `A', and therefore, source `B' seems to be younger than source `A'. V18} suggested that `A' and `B' are {young stellar objects (YSO) in their} Class II and Class I evolutionary stages, respectively. The CO outflows appear to be centred on `B' and the H$_{2}$ observations show that this is jet-driven.

OH masers were detected toward this source at 1.665 and 1.667 GHz \citep{Edris2007}. The very bright 1665 MHz peak is at (18$^{\text{h}}$17$^{\text{m}}$26$^{\text{s}}\!$.5$\pm$1$^s$.1, --17$^{\circ}$22$'$29$''\pm$16), (30, 16) arcsec {South-East (S--E)} from the IRAS position, but within the combined errors. The maser peaks at 48.33 km s$^{-1}$, only 1 km s$^{-1}$ from the deduced systemic velocity \citep{Molinari1996}. The 1665 masers are likely to be associated with the more evolved source `A' whilst the 1667 GHz line is offset $\sim$~$0'.5$ further to the E and has a more red-shifted velocity, 61.75 km s$^{-1}$. IRAS 18144--1723 is associated with  22 GHz water masers \citep{Palla1991} and 6.7 GHz methanol \citep{Caswell2009}. The 6.7 GHz transition is a Type II maser, thought to be radiatively pumped and generally found in the vicinity of  massive  protostellar objects \citep{Cragg1992}. The 44 GHz class I methanol masers imaged by \citet{GomezRuiz2016} are offset by up to $\sim20''$ from the IR source so they infer that they trace shocks due to an outflow from a deeply-embedded object.

The aim of this study is to trace 6.7 GHz methanol masers towards IRAS 18144--1723 and use  high resolution imaging of the 6.7 GHz methanol masers as a tool to characterise the high-mass star forming region IRAS 18144--1723. This paper is organised as follows: in Sec.~\ref{sec:obs} we describe observations and data reduction{, then we discuss the results we obtained in Sec.~\ref{sec:res}. Finally, we draw our conclusions in Sec.~\ref{sec:concl}}. 
\section{Observations and data reduction}
\label{sec:obs}
The high mass star forming region IRAS 18144--1723 was observed  in  C-band at 6.7 GHz on 2007 April 11 using 6 antennas of the  Multi-Element-Radio-Interferometer-Network (MERLIN; \citealt{Thomasson1986}), baselines between 11.2 -- 217 km.  Observations were carried out in dual circular polarisation and the two  polarizations were calibrated separately but all imaging was performed in total intensity; methanol maser polarisation is expected to be negligible at this spectral resolution. Table~\ref{tab:obs} summarises the main parameters of the observed sources.

The data were extracted from the MERLIN archive and converted to  FITS format at Jodrell Bank Centre for Astrophysics using the local {\tt d-programs}. The Astronomical Image Processing System (AIPS) software package was then used for the rest of the data reduction process.
We flagged bad data and calibrated the data to remove instrumental and atmospheric errors  following \citet{MUG} but applied to methanol masers (as in e.g. \citet{Darwish2020}).  The phase calibrator was observed in a wider bandwidth to maximise signal to noise, and the bandpass calibrator was used to estimate and correct the phase offset between the 16 MHz and 1 MHz bandwidth data.
The flux scale is derived from 3C286 \citep{Baars1977}, adjusted for the resolution of MERLIN giving derived flux densities for the other sources accurate to 9\%.  The final astrometric accuracy of the methanol masers (before self-calibration) is determined mainly by the accuracy of the telescope positions and  phase transfer from the phase calibrator, giving an error of 15 mas (mas refers to milli-arcsec).

The total maser bandwidth at full sensitivity is $\sim40$ km s$^{-1}$.
The brightest maser, 20 Jy in the channel at 51 km s$^{-1}$,  was used for self-calibration. The maser image cubes were produced by Fourier transformation of the calibrated data followed by deconvolution of the instrumental sampling pattern.  The spectral resolution is 0.1756 km s$^{-1}$, the off-source  noise $\sigma_{\mathrm{rms}}$ is $\sim9$ mJy and the synthesised beam is (121 $\times$ 35) mas at position angle 12$^{\circ}$.  We measured the maser positions and flux densities by fitting 2D elliptical Gaussian components to each patch of maser emission above $6 \sigma_{\mathrm{rms}}$ in each channel. We rejected isolated components or those coinciding with side lobes. Some maser spots are slightly resolved but due to the ellipticity of the synthesised beam at low elevation the beam could not be deconvolved accurately.   Components in adjacent channels with similar positions  were grouped into features. These measurements are given in Table~\ref{tab:comps}.
\begin{table*}
\centering
\caption{Measured methanol maser properties. The first column identifies the feature to which each component is assigned. Positions are offset from ($\alpha$ = 18$^{\text{h}}$17$^{\text{m}}$, $\delta$ = --17$^{\circ}$22$'$) and the errors $\sigma_{\alpha}$, $\sigma_{\delta}$ are relative (stochastic). $S$ and $\sigma_s$ are the component flux density and its error. }
\begin{tabular}{cccccccc}
\hline
F& $V_{\mathrm{LSR}}$ & $\alpha$ &$\sigma_{\alpha}$ &$\delta$ &$\sigma_{\delta}$ &$S\,\,$&$\sigma_S$\\
&(km s$^{-1}$)&(sec)&(sec)&(arcsec)&(arcsec)&(Jy)&(Jy)\\
\hline
   1&  52.31& 24.25861&   0.00010&$-$12.8069   &    0.0040&     0.19&  0.02\\
   1&  52.14& 24.25867&   0.00001&$-$12.8146   &    0.0006&     1.13&  0.02\\
   1&  51.96& 24.25869&   0.00001&$-$12.8140   &    0.0003&     2.37&  0.03\\
   1&  51.79& 24.25863&   0.00001&$-$12.8132   &    0.0004&     2.32&  0.03\\
   1&  51.61& 24.25861&   0.00001&$-$12.8044   &    0.0004&     2.29&  0.03\\
   2&  51.44& 24.25879&   0.00001&$-$12.7951   &    0.0002&     7.07&  0.04\\
   2&  51.26& 24.25891&   0.00001&$-$12.7932   &    0.0001&    16.00&  0.08\\
   2&  51.08& 24.25903&   0.00001&$-$12.7907   &    0.0001&    20.07&  0.09\\
   2&  50.91& 24.25913&   0.00001&$-$12.7885   &    0.0002&    15.14&  0.07\\
   2&  50.73& 24.25915&   0.00001&$-$12.7878   &    0.0002&     6.45&  0.04\\
   2&  50.56& 24.25897&   0.00001&$-$12.7910   &    0.0006&     1.47&  0.03\\
   2&  50.38& 24.25849&   0.00007&$-$12.8010   &    0.0027&     0.42&  0.03\\
   2&  50.21& 24.25835&   0.00010&$-$12.8156   &    0.0043&     0.26&  0.03\\
   3&  50.03& 24.25817&   0.00003&$-$12.8533   &    0.0012&     0.67&  0.03\\
   3&  49.86& 24.25811&   0.00001&$-$12.8575   &    0.0003&     4.43&  0.04\\
   3&  49.68& 24.25808&   0.00001&$-$12.8578   &    0.0002&    10.68&  0.07\\
   3&  49.50& 24.25803&   0.00001&$-$12.8590   &    0.0002&    11.28&  0.06\\
   3&  49.33& 24.25723&   0.00001&$-$12.8660   &    0.0002&     7.92&  0.05\\
   4&  49.15& 24.25608&   0.00001&$-$12.8730   &    0.0002&     7.92&  0.04\\
   4&  48.98& 24.25596&   0.00001&$-$12.8727   &    0.0002&     8.66&  0.05\\
   4&  48.80& 24.25569&   0.00001&$-$12.8732   &    0.0002&     5.74&  0.04\\
   4&  48.63& 24.25469&   0.00001&$-$12.8805   &    0.0004&     2.90&  0.03\\
   4&  48.45& 24.25389&   0.00003&$-$12.8876   &    0.0008&     1.50&  0.03\\
   4&  48.28& 24.25383&   0.00006&$-$12.8846   &    0.0018&     0.70&  0.03\\
   4&  48.10& 24.25325&   0.00024&$-$12.8732   &    0.0071&     0.20&  0.04\\
   5&  50.56& 24.24824&   0.00001&$-$12.9301   &    0.0004&     1.86&  0.03\\
   5&  50.38& 24.24823&   0.00001&$-$12.9313   &    0.0003&     3.12&  0.03\\
   5&  50.21& 24.24818&   0.00001&$-$12.9314   &    0.0004&     2.52&  0.03\\
   5&  50.03& 24.24814&   0.00001&$-$12.9310   &    0.0007&     1.27&  0.03\\
   5&  49.86& 24.24797&   0.00005&$-$12.9258   &    0.0025&     0.38&  0.03\\
   6&  49.86& 24.25052&   0.00003&$-$12.9417   &    0.0014&     0.88&  0.04\\
   6&  49.68& 24.25053&   0.00002&$-$12.9427   &    0.0009&     2.01&  0.07\\
   6&  49.50& 24.25062&   0.00002&$-$12.9421   &    0.0007&     2.33&  0.06\\
   6&  49.33& 24.25055&   0.00002&$-$12.9435   &    0.0007&     1.35&  0.04\\
   7&  47.92& 24.26411&   0.00007&$-$12.6911   &    0.0029&     0.28&  0.02\\
   7&  47.75& 24.26424&   0.00002&$-$12.6790   &    0.0008&     1.01&  0.03\\
   7&  47.57& 24.26428&   0.00001&$-$12.6760   &    0.0005&     1.94&  0.03\\
   7&  47.40& 24.26429&   0.00001&$-$12.6736   &    0.0005&     1.68&  0.03\\
   7&  47.22& 24.26429&   0.00003&$-$12.6692   &    0.0011&     0.67&  0.02\\
   7&  47.05& 24.26421&   0.00013&$-$12.6559   &    0.0057&     0.14&  0.02\\
   8&  47.40& 24.23442&   0.00006&$-$12.8118   &    0.0025&     0.38&  0.03\\
   8&  47.22& 24.23433&   0.00003&$-$12.8081   &    0.0015&     0.48&  0.02\\
   8&  47.05& 24.23426&   0.00005&$-$12.8053   &    0.0019&     0.27&  0.02\\
   8&  46.87& 24.23402&   0.00013&$-$12.8006   &    0.0055&     0.10&  0.02\\
   9&  46.17& 24.23413&   0.00011&$-$12.7984   &    0.0055&     0.11&  0.02\\
   9&  45.99& 24.23425&   0.00008&$-$12.8125   &    0.0038&     0.20&  0.02\\
   9&  45.82& 24.23425&   0.00008&$-$12.8138   &    0.0034&     0.22&  0.02\\
   9&  45.64& 24.23432&   0.00007&$-$12.8122   &    0.0031&     0.22&  0.02\\
   9&  45.47& 24.23444&   0.00007&$-$12.8160   &    0.0031&     0.23&  0.02\\
   9&  45.29& 24.23446&   0.00006&$-$12.8196   &    0.0027&     0.26&  0.02\\
   9&  45.11& 24.23444&   0.00008&$-$12.8230   &    0.0037&     0.21&  0.03\\
   9&  44.94& 24.23442&   0.00027&$-$12.8177   &    0.0115&     0.11&  0.03\\
\hline
\end{tabular}
\label{tab:comps}
\end{table*}
\section{Results and discussion}
\label{sec:res}
\begin{figure}
\includegraphics[width=\columnwidth]{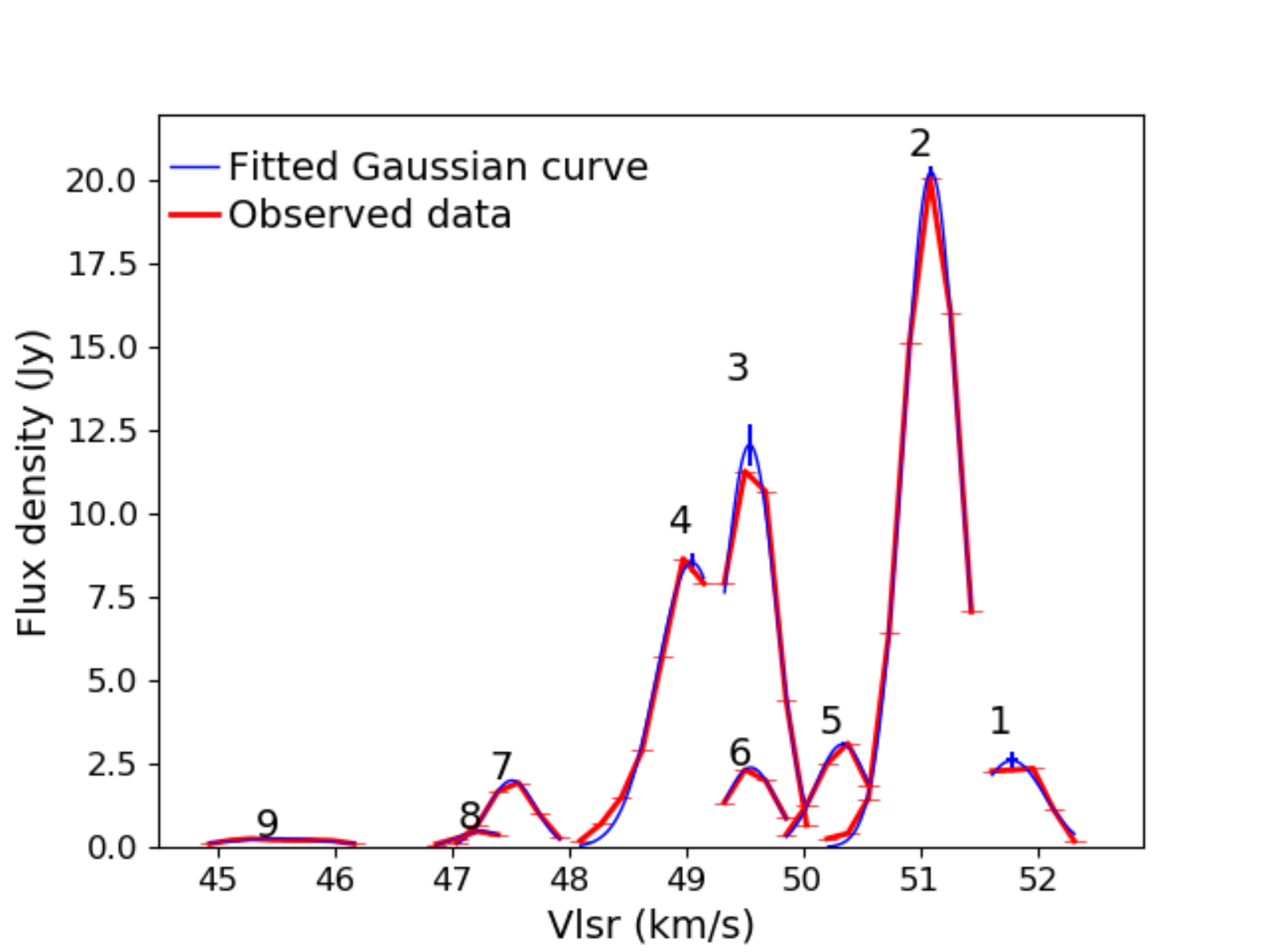}
\caption{The flux densities of the 6.7 GHz methanol maser spots in each feature (numbered) as a function of velocity are shown in red (x-errors channel width, y-errors smaller than the line width) overlaid with fitted Gaussian curves (blue) with fitting uncertainties shown. }
\label{fig-1}
\end{figure}
\begin{figure}
\includegraphics[width=\columnwidth]{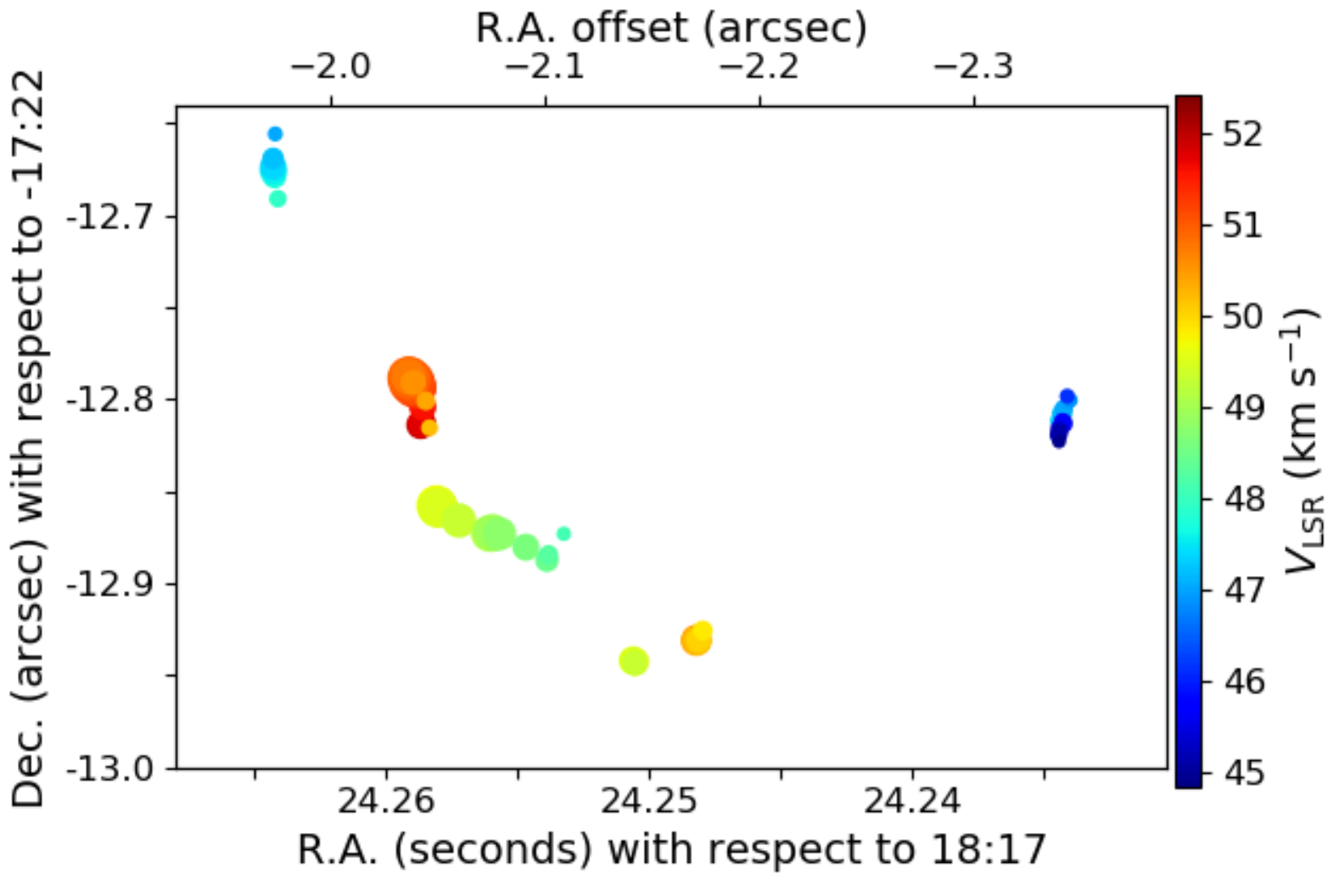}
\caption{The distribution of the 6.7 GHz methanol masers. The lower x-axis and y-axis show the right ascension and declination. The upper x-axis is the right ascension offset in arcsec from the pointing position (18$^{\text{h}}$17$^{\text{m}}$24$^{\text{s}}\!$.40, --17$^{\circ}$22$'$13\farcs0). The colours represent velocity and the symbol area is proportional to flux density. The relative position uncertainties are similar to or less than the spot sizes. }
\label{fig-2}
\end{figure}
\subsection{Comparing locations of methanol masers and IR sources in IRAS 18144--1723}
\label{sec:res:one}
\begin{figure}
  \includegraphics[width=\columnwidth]{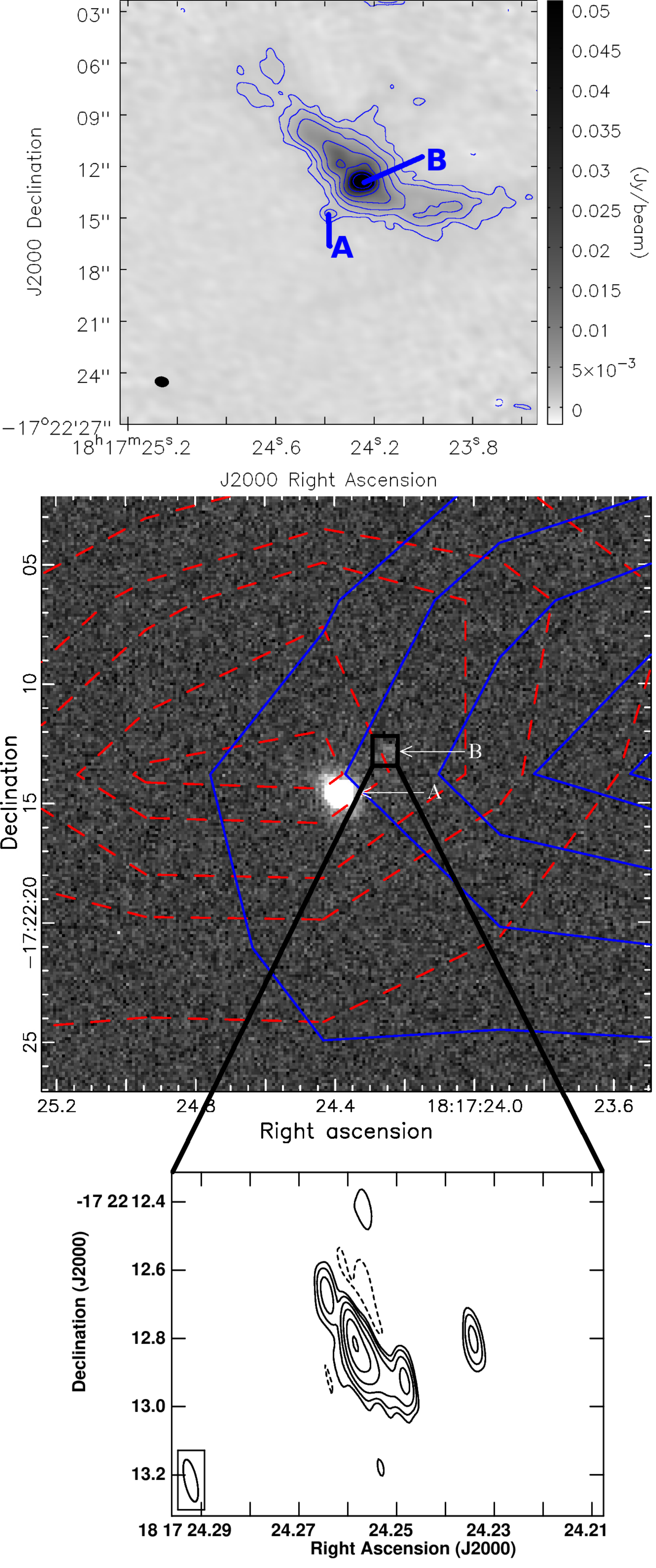}
 \vspace*{-0.2cm}
 \caption{{Middle panel: IR sources `A' and `B' detected by V18 (their fig. 2). The red and blue contours are generated from integrated CO J=3-2 emission of the receding and approaching outflow lobes. Top panel: Same area of ALMA continuum observations at $\lambda$ 1.3 mm, contours at (1,2,4...)$\times$ 1 mJy (5 $\sigma_{\mathrm{rms}}$), restoring beam shown lower left (courtesy A. Avison, see Sec.~\ref{sec:res:one}). Bottom panel: zoom in to the integrated 6.7 GHz methanol maser emission towards IRAS 18144--1723 imaged using MERLIN. The contours are at (--1, 1, 2, 4...)$\times$ 19 mJy beam$^{-1}$, restoring beam  shown lower left. It shows that the 6.7 GHz methanol maser position is consistent with that of the weaker IR source, `B'.}}
\label{fig-3}
\end{figure}

\begin{figure}
\includegraphics[width=\columnwidth]{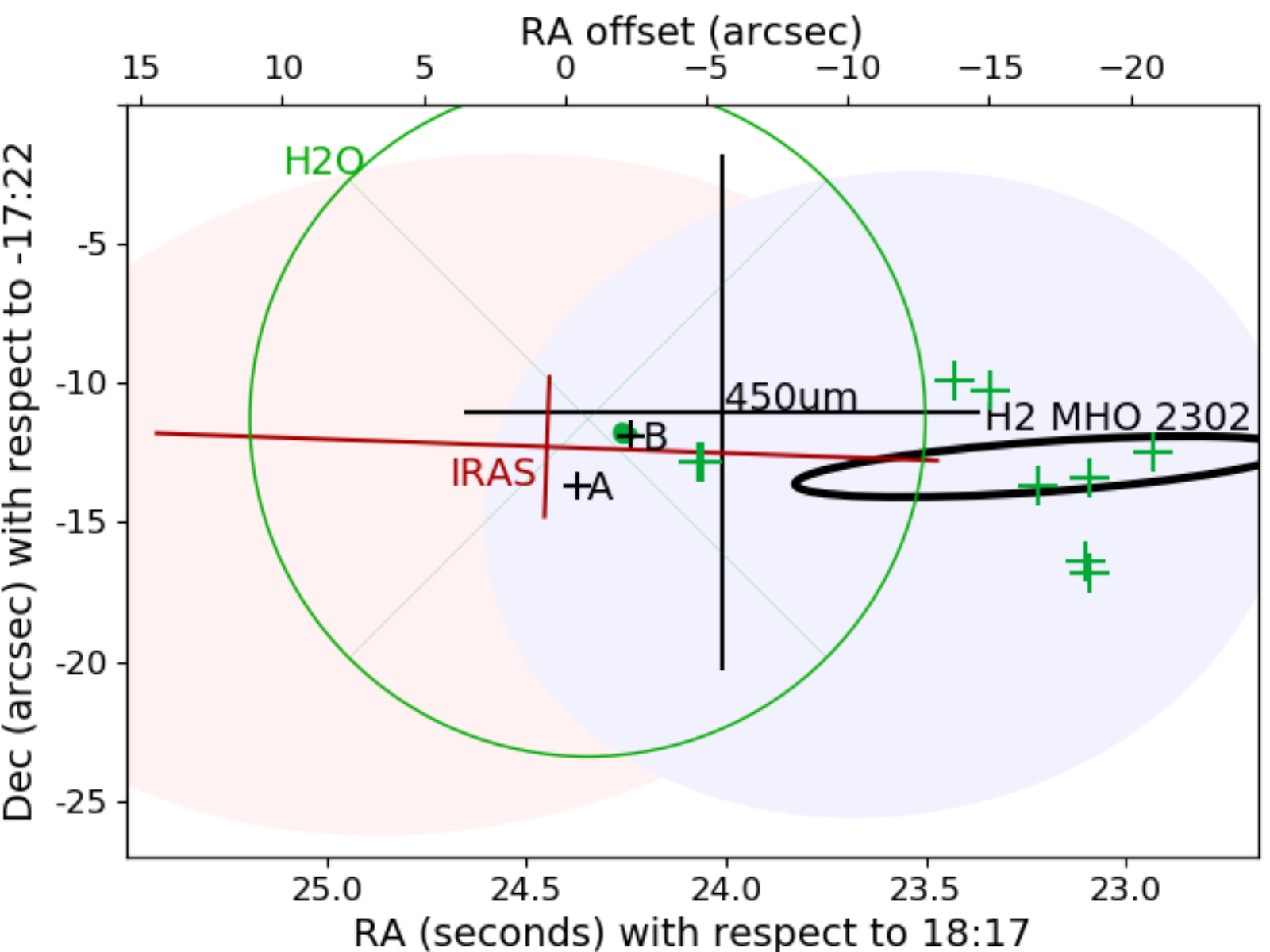}
\caption{Overview of the relative locations and kinematics of tracers of IRAS 18144--1723. The lower x-axis and y-axis show the right ascension and declination; the upper x-axis shows the arcsec offset from $\alpha$ (18$^{\text{h}}$17$^{\text{m}}$24$^{\text{s}}\!$.40). The background blue and red shaded ellipses on the right and left show CO emission in the velocity ranges  23.3 to 44.0 and 50.6 to 71.3 km s$^{-1}$, respectively (V18). The black ellipse represents the extent of the outflow traced in H$_2$, from \citet{Varricatt2010}. The large dark red cross shows the IRAS position \citep{IRAS1988}, the tilted error bars representing the error ellipse. The large black cross is the 450 micron wavelength SCUBA-2 peak and position errors  and the small crosses labelled `A' and `B' are the peaks identified from UIST M'-band (4.69 micron) observations (V18).  The large green circle is the pointing accuracy for 22 GHz water maser peaks at 45.35 and 51.32 km s$^{-1}$, \citep{Palla1991}. The green crosses are 44 GHz methanol masers, assuming 1$"$ position accuracy \citep{GomezRuiz2016}, velocities 46.6 to 48.3 km s$^{-1}$.  The green dot at `B' marks the positions of 6.7 GHz masers (this work), see Table~\ref{tab:comps}. }
\label{fig-4}
\end{figure}

We found  fifty two 6.7 GHz methanol maser spots in the velocity range 44.9- - 52.3 km s$^{-1}$, grouped into 9 features, as labelled in the spectrum, shown in Fig.~\ref{fig-1}, all of which have close to Gaussian spectral profiles. The distribution of the maser components is shown in Fig.~\ref{fig-2} while Fig.~\ref{fig-3} (bottom and middle panels) shows a contour map of the maser emission integrated over all channels compared with figure 2 from V18 which shows that the 6.7 GHz masers lie within the 0\farcs5 uncertainty in the position of source `B'.
As seen from Fig.~\ref{fig-4} the 6.7 GHz masers form a compact clump centred within the position uncertainty of the IRAS source, while the 44 GHz masers are spread along an E--W axis in the direction of the CO outflow. For further details see \citet{GomezRuiz2016}, but note that paper was prior to the discovery of IRAS 18144--1723 `B'.

Figure~\ref{fig-4} shows that the 6.7 GHz methanol masers lie within the position uncertainty of source `B'. Their velocity range (44.9 -- 52.3 km s$^{-1}$) covers the systemic velocity of 47.3 km s$^{-1}$ \citep{Molinari1996} and the velocities of the water maser peaks, but is only 15\% of the CO velocity range. In contrast, the 44 GHz masers appear to be associated with the blue-shifted wing of the outflow, located up to $\sim20''$ from `B', much more remote than the 6.7 GHz masers. This is consistent with the properties of Class II methanol masers which are collisionally excited, typically by the interaction of outflows with the interstellar medium (ISM). Class II methanol masers are confined to velocities 46 -- 49 km s$^{-1}$, a smaller radial velocity range than the 6.7 GHz masers and far slower (with respect to the systemic velocity) than the  CO seen in the same direction, which \citet{GomezRuiz2016} interpret as showing that the 44 GHz masers arise from the ISM after being shocked by interaction with the outflow from IRAS 18144--1723.
\begin{figure}
\includegraphics[width=\columnwidth]{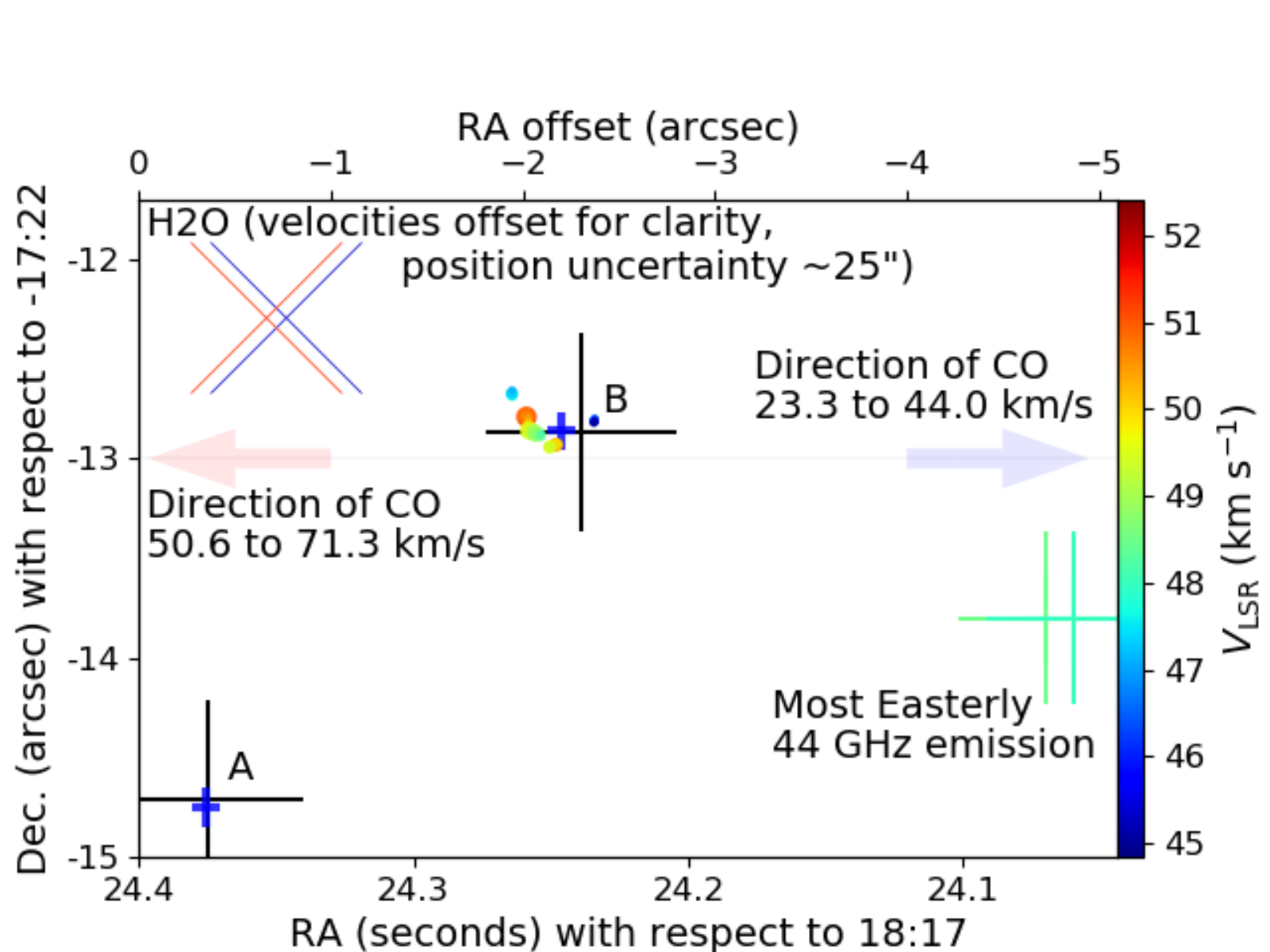}
\caption {The inner few arcsec around IRAS 18144--1723. The lower x-axis and y-axis show the right ascension and declination; the upper x-axis shows the arcsec offset from $\alpha$  (18$^{\text{h}}$17$^{\text{m}}$24$^{\text{s}}\!$.40). {The blue plus signs mark the peaks of the ALMA $\lambda$ 1.3 continuum.} The coloured spots are the 6.7 GHz masers (see Fig.~\ref{fig-2}) and other species are as labelled; see Fig.~\ref{fig-4}.}
\label{fig-5}
\end{figure}
Fig.~\ref{fig-5} shows the large offset ($\sim$ 0.095 pc) between the closest 44 GHz and 6.7 GHz masers. The single-dish water maser position uncertainty is far greater, but their velocity span  lies within 1 km s$^{-1}$ of that of the 6.7 GHz masers. The position uncertainty for source `B' means it is not clear whether the 6.7 GHz masers (total extent $\sim$ {1982 au}) are spatially centred on the IR source and the position-velocity relationship is not clearly either a disc or an outflow. Therefore, we investigate fitting models for a Keplerian disc and a biconical outflow.

IRAS 18144--1723 is one of the targets in two ALMA proposals; project ID 2015.1.01312.S and the ongoing project `ALMAGal' (ID: 2019.1.00195.L) for which observations are still partly proprietary. Archive image cubes for  project ID 2015.1.01312.S show that the arcsec-scale velocity gradients in the region are complex. Hence, it is not obvious how to use these data to elucidate the morphology of the methanol 6.7 GHz masers, since their whole extent fits within one mm-wave resolution element of  0\farcs824 $\times$ 0\farcs602 at position angle $81^{\circ}$. {A preview of the mm-wave continuum (taken from Avison and Fuller, private communication) is given, in this study, in the top panel of Fig.~\ref{fig-3}. The illustration} shows that the emission is dominated by a bright peak. We fitted 2-D Gaussian components to measure two peaks of $2.83\pm0.07$ and $160\pm6$ mJy, deconvolved FWHM $\sim$ 350 and $\sim$ 315 mas, {at positions ($\alpha$, $\delta$) of (18$^{\text{h}}$17$^{\text{m}}$24$^{\text{s}}\!$.3757, --17$^{\circ}$22$'$14\farcs750) and (18$^{\text{h}}$17$^{\text{m}}$24$^{\text{s}}\!$.2465, --17$^{\circ}$22$'$12\farcs861), respectively}. Allowing for stochastic and astrometric errors, the {position uncertainties are about (100, 75) and (95, 70) mas, respectively,} which allows these peaks to be identified with V18 sources  `A' and `B', which have 0\farcs5 position uncertainties. The ALMA positions are marked on Fig.~\ref{fig-5}. However, Fig.~\ref{fig-3}, top, also shows  much diffuse, arcsec-scale mm-wave emission at different position angles.
These ALMA data suggest that `B' has a larger concentration of dust than `A', supporting the younger age of `B'. Whilst the Westerly dust extension is in a similar direction to the blue-shifted CO outflow, the N--E extension is misaligned with the red-shifted CO wing. A fuller analysis will give more insights (Avison et al., in prep).
\subsection{Fitting potential Keplerian disc}
\label{fit-kep}
We investigated whether the distribution of the 6.7 GHz masers was consistent with a model of Keplerian motion around source `B' for a range of masses for the central object.  We measured the positions of the maser components along an axis (representing the plane of the disc) projected onto the sky at orientations to the North-South (N-S) axis at 35 degrees to the East.
We took the systemic velocity ($V_{\rm c}$) as 47.3 km s$^{-1}$ \citep{Molinari1996}. We  used the model described in \citet{Edris2005} to  generate position-velocity curves for various masses of the central object.

We ignored the inclination, $i$, of the disc axis to the line of sight; V18 estimate that this is $\sim32^{\circ}$, possibly up to  $\sim57^{\circ}$.  This means that the disc itself is tilted at between $33^{\circ}-58^{\circ}$ to the line of sight so the $V_{\mathrm{LSR}}$ offsets we measure from the systemic velocity are  scaled by $\sin~{i}$, 15--45\% less than if the disc was edge on. As discussed in \citet{Seifried2016}, this by itself does not affect the shape of a Keplerian position-velocity diagram but would lead to an underestimate of the enclosed mass if a fit was possible.
\citet{Seifried2016} discuss modelling thermal molecular emission from Keplerian protostellar discs. They find that the fit can be quite accurate for $i>30^{\circ}$. However, even if the gas traces spirals and infall, such that the position-velocity relationship is not smooth, the overall trend should show a decrease of velocity with increasing radius, which is not seen in these methanol masers.

\begin{figure}
\includegraphics[width=\columnwidth]{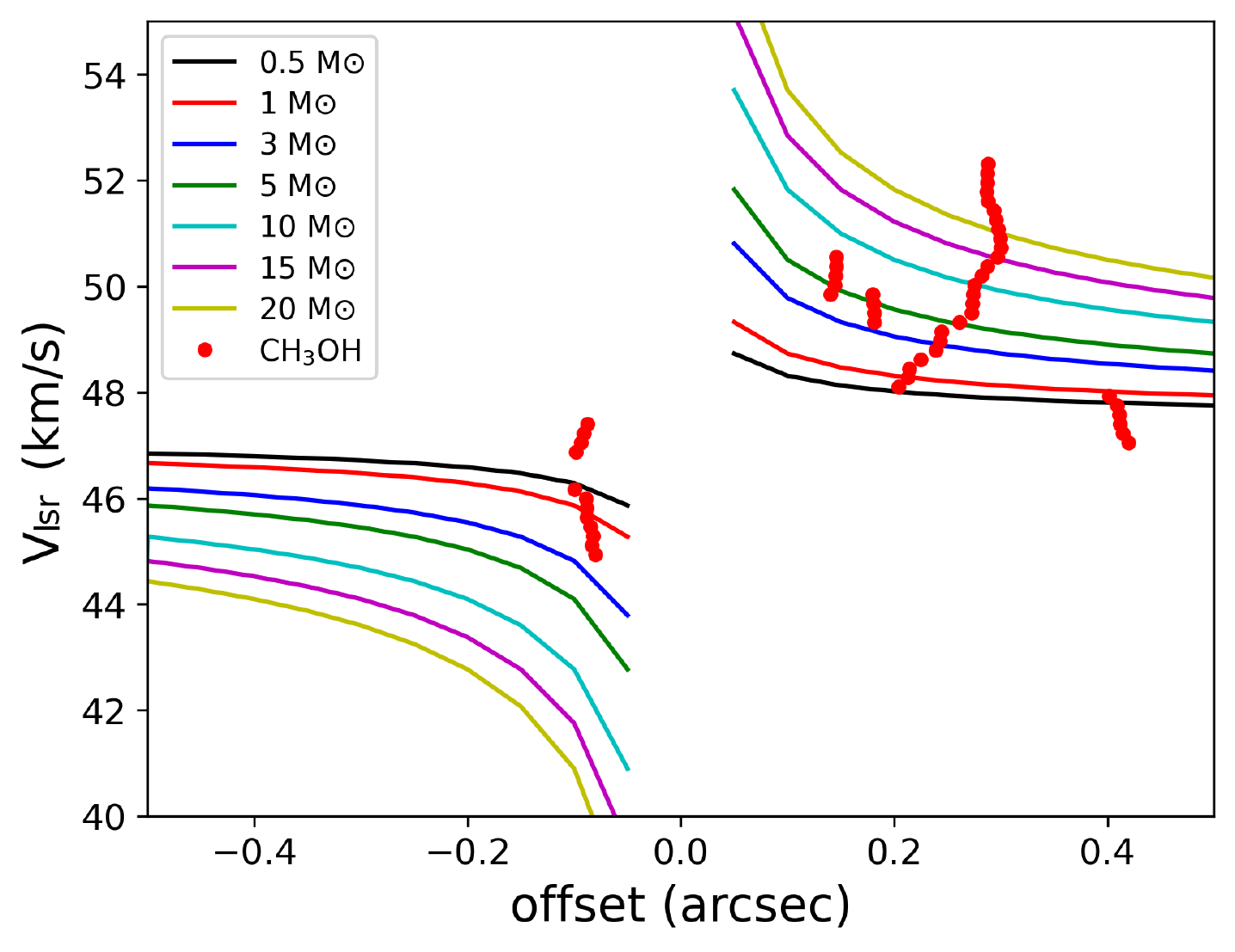}
\caption{Position-velocity diagram for the CH$_3$OH maser spots (red circles). The x-axis represents the offset of the maser spots with respect to the centre of mass if they were in a rotating disc. The angular separations were measured in the East to West direction, with (0) at $\alpha$ = 18$^{\text{h}}$17$^{\text{m}}$24$^{\text{s}}\!$.239 and $\delta$ = --17$^{\circ}$22$'$12\farcs87.  The coloured curves indicate the Keplerian motion velocities around different central masses at a systemic velocity of 47.3 km s$^{-1}$; {see figure key}.}
\label{fig-6}
\end{figure}

In order to identify a Keplerian disc, the observed maser velocity -- position relationship should be consistent with a single central mass.
Fig.~\ref{fig-6} compares the observed distribution of spots with the Keplerian model for a disc rotating about an E-W axis in the direction of the CO jets. The maser spot position-velocity distribution is not consistent with any specific stellar mass.  We also tested other disc orientations but did not obtain any reasonable fits. The position-velocity diagrams for these are given in Appendix~\ref{sec_Kepler4}.

It appears from these comparisons that there is no clear relationship between the distribution of the 6.7 GHz masers and the proposed Keplerian disc. This opens the door to other possibilities for the cause of the methanol maser distribution and kinematics, including the possibility of several enclosed sources.
\subsection{Testing models for bipolar outflows or a disc and the implications of mas-scale structures}
\label{outflow}
We tested the kinematic model of a bipolar outflow by \citet{Moscadelli2000}, as applied to the {single-epoch observations} of 6.7 GHz methanol masers by \citet{Bartkiewicz2009}. {The model assumes that the masers arise on the surface of a conical bipolar jet and the vertex of the cone coincides with the central object(s). The velocity of a spot is constant and is directed radially outward from the vertex. The vertex is also taken as the centre of the coordinate system, so that the z-axis is along the line-of-sight and the x-axis coincides with the projection of the jet axis on the plane of the sky. Similarly as in Sec. \ref{fit-kep}, we took the systemic velocity ($V_{\rm c}$) as 47.3 km s$^{-1}$ and {minimized the $\chi^2$ function for the 52 detected maser spots with their observed V$_{\mathrm{LSR}}^j$ (listed in Table~\ref{tab:comps}) following the expression of Eq.~(3) in} \citet{Moscadelli2000}:
\begin{eqnarray}
{\rm \chi^2= \sum\limits_{j=2}^{52} \bigg(\frac{V_z^j}{V_z^1}-\frac{V_{\mathrm{LSR}}^j-V_c}
{V_{\mathrm{LSR}}^1-V_c}\bigg)}^2, \nonumber
\label{eq:chisq}
\end{eqnarray}
The V$_{\rm z}^{\rm j}$ is the velocity of the j$^{th}$ spot calculated using Eqs. (1) and (2) of \citet{Moscadelli2000}. Note, the erratum of these equations in \citet{Moscadelli2005}.}

{For the location of the origin, we took ranges in {$\alpha$ and $\delta$} of ($-$4" to $-$1") with respect to {($\alpha$ = 18$^{\text{h}}$17$^{\text{m}}$24$^{\text{s}}\!$.4)} and ($-$12" to $-$14") with respect to  {($\delta$ = --17$^{\circ}$22$'$)}} covering the position uncertainty of source `B', see Fig.~\ref{fig-5}. {We took samples along both of these coordinates with} a step size of 20~mas. {The axis of the outflow (i.e. the model x-axis) was sampled in a step size of 5$^{\rm o}$, within an angular range from $-$45$^{\rm o}$ to $+$45$^{\rm o}$  with respect to the E--W direction}, as constrained by the CO outflow. The inclination angle, $\Psi$, between the outflow axis  (i.e. the z-axis) and the line of sight, and the opening angle of the outflow, 2$\theta$,  were set to ranges from 0$^{\rm o}$ to 180$^{\rm o}$ and sampled with a step of 5$^{\rm o}$.
\begin{figure*}
\includegraphics[scale=0.6]{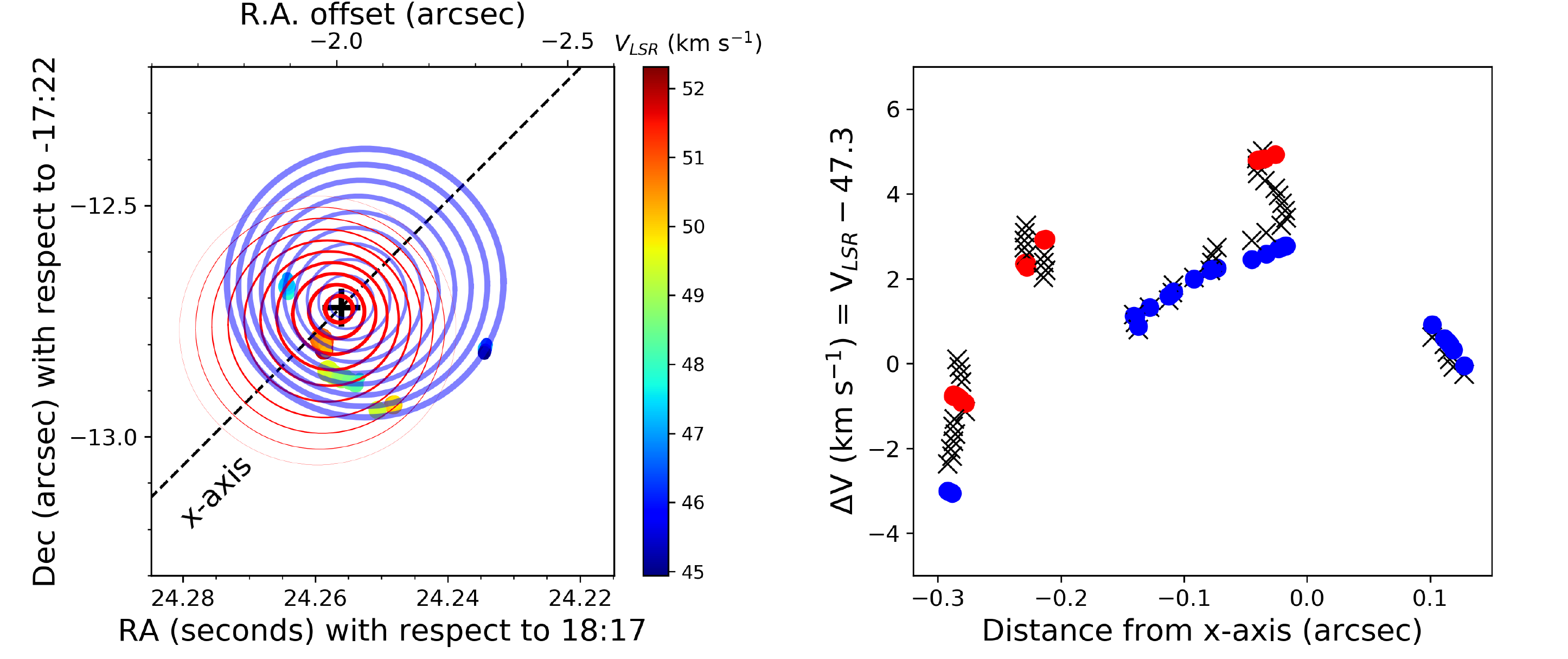}
\caption{Outflow model fitted to 6.7~GHz methanol masers in IRAS 18144 according to the model of \citet{Moscadelli2000}.  Left panel: The distribution of masers as in Fig.~\ref{fig-2}.
The dashed line and the cross trace the fitted outflow axis (x-axis) and the vertices of the cones, respectively. The opening angle of the fitted cone of 172$^{\circ}$ indicates that the outflow covers the whole area. The overlay shows cross-sections of the fitted bicone in blue and red for the approaching and receding outflows. Right panel: Comparison of observed data (red and blue symbols, for components red- and blue-shifted with respect to $V_{\mathrm C}$) and the model data (x symbols).}
\label{fig-7}
\end{figure*}
\begin{figure*}
\includegraphics[width=18 cm]{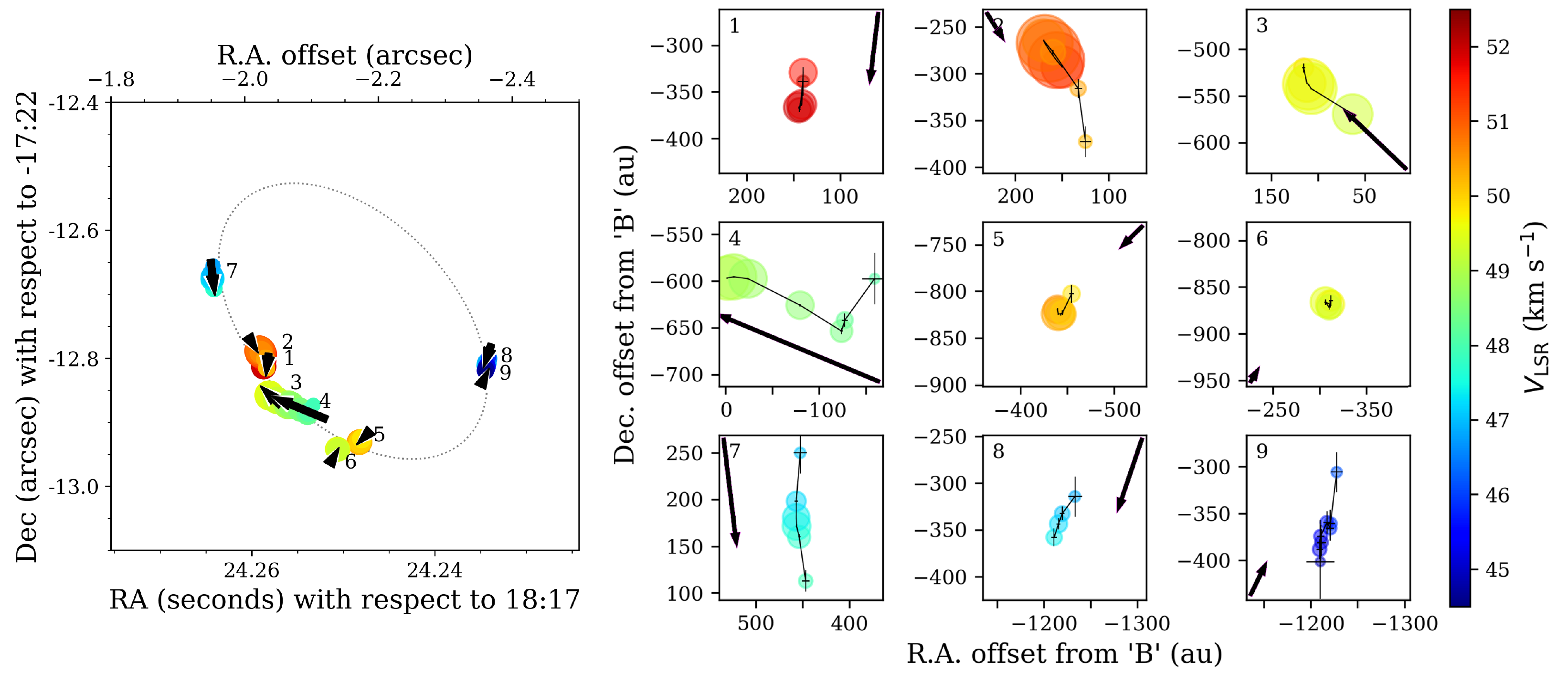}
\caption{The orientation of velocity gradients in each group of 6.7 GHz methanol masers, {as located on the sky plane in the left panel and individually in the right panel. Features are} numbered according to the spectral features in Fig.~\ref{fig-1} and Table~\ref{tab:comps}. The arrows trace the trends from the blue- to red-shifted velocities of spots within each group, {i.e. the vector (d$\alpha$/d$V_{\mathrm{LSR}}$, d$\delta$/d$V_{\mathrm{LSR}}$)
 calculated as described in Sec. \ref{Vgrad}.}}
\label{fig-8}
\end{figure*}
The smallest value of $\chi^2$ was 0.69, obtained for the following angles: $\Psi$=166$^{\rm o}$, 2$\theta$=172$^{\rm o}$ and an x-axis at a position angle of 45$^{\circ}$. The fitted outflow was centred on (18$^{\text{h}}$17$^{\text{m}}$24$^{\text{s}}\!$.256, --17$^{\circ}$22$'$12\farcs72), i.e. ($\alpha$ offset,$\delta$ offset) = (--2.06",0.28") from the position of the IRAS source. Fig.~\ref{fig-7} shows the best-fitting orientation of the x-axis in projection; the high value of $\Psi$ means that the outflow is oriented almost directly towards the observer, approaching us in the north-west. Similar to the findings of \citet{Bartkiewicz2016}, we note that for this target, the opening angle of the fitted outflow is large, and the outflow is not well-collimated, but covers the whole area where the 6.7 GHz masers are seen. Such apparent behaviour could be a characteristic of outflows  related to methanol maser emission. This was indicated by \citet{Seifried2012} in their numerical simulations of outflows formed during the collapse of 100 M$\odot$ cloud cores; the outflow can have a sphere-like morphology at an early stage of evolution, expanding with the velocity up to a few km s$^{-1}$ in all directions (their fig. 7). Furthermore, we note  the misalignment of the E--W CO red- and blue-shifted wings (extending for arcmin) with respect to the SE-NW projection of the axis of outflow  fitted to the methanol masers on the sub-arcsecond scale.

In order to analyse the distribution of the methanol masers further, we note their distribution can be classified as arc or ring-like according to \citet{Bartkiewicz2016}. In Fig. ~\ref{fig-8}, left panel we show the best-fitted ellipse to the maser spots using the code by \citet{f99}. However, the  internal velocity gradients within each maser feature  do not show any overall trend consistent with {simple} rotation or expansion/infall, but instead  change direction from one feature to another without regularities. {In some cases the gradients are  parallel to the fitted ellipse; in others they are perpendicular,  oriented towards or away from the centre} (see also Fig.~\ref{fig-8}). We also {point out that} the major axis is perpendicular to the {fitted} outflow axis ({compare with Fig.~\ref{fig-7}}) {and thus is misaligned with the East-West CO outflow.}

However, the most red- and blue-shifted maser features (labelled as 1, 2, and 9) lie along the axis of the CO outflow, which could be disrupting or sweeping up the maser-bearing region. \citet{Torstensson2011} proposed that the methanol masers in well-known HMSFR Cepheus A HW2 appeared in  shocked regions between the outflowing or infalling gas, at the interface with the accretion disk. In that case, both kinematical regimes will influence the masers. If there is a disc present but it is misaligned with the CO outflow, source `B' may be multiple. The observations of V18 were at a resolution $\ge$ 0\farcs12, so IR sources separated by less than 470 au would not have been distinguished. Alternatively, the CO outflow may change direction;  high-angular resolution studies of CO would be needed to investigate whether the 6.7 GHz methanol masers really indicate small-scale motions that differ from the large-scale outflows. Proper motions of the methanol masers, measured using multiple observations over several years, at mas resolution, would  constrain the disc or outflow models.
\subsection{Gradients of $V_{\mathrm{LSR}}$}
\label{Vgrad}
We have a single epoch of maser observations (so far) and thus cannot calculate 3D velocities but we can investigate the relationship between the line-of-sight velocity ($V_{{\mathrm{LSR}}}$) and the projected positions of the  6.7 GHz methanol masers. We calculated the $V_{\mathrm{LSR}}$ gradient within each of the 9 maser clumps as a function of position change in the right ascension and declination directions, $\alpha$ and $\delta$, i.e. $\mathrm{d}V_{\mathrm{LSR}}/\mathrm{d}\alpha$, $\mathrm{d}V_{\mathrm{LSR}}/\mathrm{d}\delta$. For convenience, we took the nominal centre of the maser distribution found in Sec. \ref{outflow}, (18$^{\text{h}}$17$^{\text{m}}$24$^{\text{s}}$.256, --17$^{\circ}$22$'$12\farcs72), as the coordinate origin in the $\alpha\delta$ plane and converted angular separations to au (using a distance of 3.9 kpc) as convenient for physical visualisation. The gradients within each clump seem to follow monotonic curves, apart from the weakest maser spots, which implies that deviations from a linear gradient are mainly due to turbulence or other physical effects, not measurement errors. Therefore, we used a non-weighted fit to the maser components with S/N $>18$ and used the position dispersions to estimate the errors. We also calculated the inverse; i.e. the change of position with  $V_{\mathrm{LSR}}$ ($\mathrm{d}\alpha/\mathrm{d}V_{\mathrm{LSR}}$  $\mathrm{d}\delta/\mathrm{d}V_{\mathrm{LSR}}$). 
The gradients for each feature are given in Table~\ref{tab:Vlsr-pos}.
\begin{table*}
  \caption{The internal gradients for each feature. The angular displacements in {$\alpha$ and $\delta$} as a function of $V_{\mathrm{LSR}}$, (d$\alpha$/d$V_{\mathrm{LSR}}$, d$\delta$/d$V_{\mathrm{LSR}}$) and uncertainties, were calculated as described in {Sec.~\ref{Vgrad}}, and also the inverse, i.e. gradients of   $V_{\mathrm{LSR}}$ with $\alpha$ and $\delta$.}
  \begin{tabular}{lcccccccc}
    \hline
    F&d$\alpha$/d$V_{\mathrm{LSR}}$& $\sigma_{\mathrm{d}\alpha/\mathrm{d}V_{\mathrm{LSR}}}$& d$\delta$/d$V_{\mathrm{LSR}}$& $\sigma_{\mathrm{d}\delta/\mathrm{d}V_{\mathrm{LSR}}}$&d$V_{\mathrm{LSR}}/$d$\alpha$&$\sigma_{\mathrm{d}V_{\mathrm{LSR}}/\mathrm{d}\alpha}$&d$V_{\mathrm{LSR}}/$d$\delta$&$\sigma_{\mathrm{d}V_{\mathrm{LSR}}/\mathrm{d}\alpha}$\\
   &\multicolumn{4}{c}{(au (km s$^{-1}$)$^{-1}$)}& \multicolumn{4}{c}{(km s$^{-1}$ au$^{-1}$)}\\
\hline
  1&    8.45&    3.04& --69.80&   31.35&    0.1183&  0.0426&--0.0143&  0.0064\\
  2& --15.49&    8.59& --24.69&   10.82&  --0.0646&  0.0358&--0.0405&  0.0178\\
  3&   62.69&   27.58&   59.78&   14.29&    0.0160&  0.0070&  0.0167&  0.0040\\
  4&  168.05&   23.53&   69.87&   17.43&    0.0060&  0.0008&  0.0143&  0.0036\\
  5&   19.98&    5.34& --19.76&   15.36&    0.0500&  0.0134&--0.0506&  0.0394\\
  6&  --6.21&    6.81&   10.67&    5.82&  --0.1610&  0.1765&  0.0937&  0.0511\\
  7& --13.33&    4.74&--109.31&   22.66&  --0.0750&  0.0266&--0.0091&  0.0019\\
  8&   24.45&    1.24& --72.23&    5.65&    0.0409&  0.0021&--0.0138&  0.0011\\
  9& --14.53&    3.41&   29.96&    9.43&  --0.0688&  0.0161&  0.0334&  0.0105\\
  \hline
  \end{tabular}
  \label{tab:Vlsr-pos}
\end{table*}
Fig. ~\ref{fig-8} left panel shows shows that the overall maser distribution is compatible with a tilted disc but the gradients of position change with $V_{{\mathrm{LSR}}}$  are not consistent with rotation nor with a simple bipolar outflow. This is expanded in Fig.~\ref{fig-8}, right panel, where the location of each arrow is purely for clarity, but the length and direction represents the magnitude and direction of position change with $V_{{\mathrm{LSR}}}$.  The non-systematic variation in gradients around the disc  indicates that these masers may shows an interaction between the disc and an outflow, leading to individual maser clumps being swept in different directions or affected by turbulence.

We investigated the implications of the gradients of $V_{{\mathrm{LSR}}}$ with $\alpha$ and $\delta$ using a variance-covariance matrix as described by \citet{Bloemhof20}. In Equation~\ref{eq:matrix} the diagonal elements are the variances over all features of the gradients $\mathrm{d}\alpha/\mathrm{d}V_{\mathrm{LSR}}$,
  $\mathrm{d}\delta/\mathrm{d}V_{\mathrm{LSR}}$ and the off-diagonal elements are their covariances. The units are (km s$^{-1}$ au$^{-1}$)$^{2}$. As explained by \citet{Bloemhof20}, this provides an indication of the direction of maximum velocity gradient dispersion which is independent from the origin of coordinates, i.e. it does not matter if the position of the exciting source is unknown.
\begin{equation}
\begin{bmatrix}
0.007099 & -0.002258 \\
-0.002258 &  0.001870
\end{bmatrix}
\label{eq:matrix}
\end{equation}

Diagonalising the matrix gave the eigenvector $[\lambda_{\alpha}, \lambda_{\delta}] = [0.00103, 0.00794]$ suggesting that the maximum rate of change of velocity with position is in directions within $\arctan{(\lambda_{\alpha}/\lambda_{\delta})}\sim$ 7$\pm$6$^{\circ}$ of the $\alpha$-axis (E-W), and is positive (increasingly red-shifted) in the direction close to positive $\alpha$ (W). This is also close to the CO outflow direction. The uncertainty is based on the average fractional measurement errors in ($\mathrm{d}V_{\mathrm{LSR}}/\mathrm{d}\alpha$) and ($\mathrm{d}V_{\mathrm{LSR}}/\mathrm{d}\delta$), as given in Table \ref{tab:Vlsr-pos}, but, as noted above, the biggest uncertainty is probably due to intrinsic irregularities in the maser motions.

Alternatively, in a Keplerian disc, the total variance of $dV_{\mathrm{LSR}}$ with position is zero in the direction perpendicular to the axis about which the disc is rotating. The variance is greatest at position angles with respect to the rotation axis of around roughly $60^{\circ}$. This implies a rotation axis at a position angle of about $(10\pm60)^{\circ}$ with respect to the $\delta$-axis. This is at odds  with the NW-SE axis of rotation of the disc fitted in Sections \ref{fit-kep} and~\ref{outflow}. Moreover, the variance study is based on the magnitude but not the direction of velocity gradients within clumps and inspection of Fig.~\ref{fig-8} shows reversals of direction incompatible with simple rotation.
\section{Conclusions}
\label{sec:concl}
We detected 9 maser features in the 6.7 GHz transition of methanol in the region of IRAS 18144--1723, with a peak of 20 Jy at 51.1 km s$^{-1}$. Throughout this work, we adopted a distance of 3.9 kpc to the source; on which we base all conversions from angular to linear units. The masers cover a $V_{\mathrm{LSR}}$ range of 45 -- 52 km s$^{-1}$, spanning the systemic velocity and the middle of the CO outflow (V18) which covers a velocity range 23 -- 71 km s$^{-1}$. The maser angular extent is $\sim$ 0\farcs5 ($\sim$ 1982 au at 3.9 kpc) with an astrometric accuracy of 15 mas and they are located within the position uncertainty of IR source `B'. Source `B' is only detected at wavelengths $\ge4.69$ $\mu$m and was identified by V18 as a high mass young protostar in an early stage of evolution, the origin of the E--W CO outflow. The CO velocity gradient and their model of the dusty protostellar disc suggest an axis at an angle of inclination of $32^{\circ} - 57^{\circ}$ to the line of sight, approaching in the West.  `B' is also the brightest mm-wave continuum source in the vicinity of  IRAS 18144--1723 (Avison, private communication), supporting its nature as a young Class I protostar.

It seems likely that the closest-detected OH masers \citep{Edris2007} and UCH{\sc ii} region \citep{Molinari1998} are offset by more than the astrometric errors and not associated with  source `B'. The chronology in fig. 6 of \citet{Breen2010} suggests that the association of 6.7 GHz and 44 GHz methanol masers and water masers, but not with OH nor radio continuum, places source `B' within an evolutionary stage 10000 -- 20000 yrs after the first tracers of high mass star formation appear. The accurate alignment of `B' with 6.7 GHz masers confirms its high mass nature as 6.7 GHz methanol masers are exclusively associated with high-mass protostars \citep{Breen2013}.  Observations by \citet{Caswell2009} in 2005 at $\sim1''$ accuracy recorded a similar velocity range of 47--53 km s$^{-1}$ and a peak of 32 Jy at 51.2 km s$^{-1}$, suggesting the distribution is stable, but the source has dimmed somewhat.

The 6.7 GHz masers appear to trace a patchy ellipse but the velocity distribution is not consistent with a Keplerian disc at any orientation. The maser clump internal velocities have no consistent trend around the ellipse as a whole, and even show different position-velocity gradients in adjacent clumps. This is incompatible with pure rotation, even if modified by acretion (e.g. as discussed by \citealt{Seifried2016}).

Neither do the masers trace a simple, well-collimated outflow, as although the most extreme blue-shifted emission is East of the most red-shifted emission, the maser distribution follows an irregular arc and the position-velocity gradient is not monotonic. Another possibility is a biconical outflow with a very wide opening angle, as suggested for some other sources (e.g. \citealt{Bartkiewicz2016}, and \citealt{Seifried2012}), but the best fit model implies an axis oriented approximately NW-SE at an angle of inclination within 10$^{\circ}$ of the line of sight, inconsistent with the CO outflow axis.

Matrix analysis of the gradients of $V_{\mathrm{LSR}}$ with position, following \citet{Bloemhof20}, indicates the highest velocity gradient dispersion is along a position angle of magnitude $\sim10^{\circ}$ with respect to the E-W axis, close to the CO outflow direction. Thus, the analytic models we have considered  and the matrix analysis suggest that if the 6.7 GHz masers are associated with a disc, this is disrupted or interacting with an outflow. Although pre-main-sequence stellar mergers are probably very rare, the aftermath of a star+star+disc encounter  could leave a disrupted disc (\citealt{Rivilla2013}, their appendix A.1.2). Non-merger interactions (fly-by's) are commoner and would also cause asymmetric distortion.
The complex kinematics  could also be influenced by the infall of matter and transfer of angular momentum along filaments \citep{Kong2019}. Another complication is that the orientation of such a disc (or biconical outflow) suggests an axis approximately SE-NW, or even closer to N-S. However, on arcsecond scales, the CO outflow is oriented E-W.

There could be multiple protostars within the region, which may be revealed by the high-resolution ALMA observations. If there is a single, dominant exciting source then the likeliest explanation for the 6.7 GHz maser distribution is an interaction between an outflow and a disc. Fig.~\ref{fig-8} shows that an ellipse representing a tilted disc can be fitted to the overall distribution of the clumps and the velocity inconsistencies can result from not only material directly swept up but also the ensuing turbulence and disruption of the disc.
The extreme red-shifted 6.7 GHz methanol maser clumps (1 and 2) and blue-shifted (8 and 9) are offset East and West in a similar direction as the CO outflow but the disc axis seems misaligned by about 45$^{\circ}$; however, we note that the H$_2$ outflow Fig. \ref {fig-4} has an intermediate position angle and the outflow may be flared or change direction.
\section*{Acknowledgements}
MERLIN was the UK National Radio Astronomy facility, funded by STFC, now replaced by e-MERLIN (www.e-merlin.ac.uk).
I.K is grateful to Dr. Mohamed Darwish (NRIAG - Egypt) for helping to produce the illustrations that appeared in this work. We thank Dr. A. Avison and Prof. G. Fuller for information about the ALMA data. We thank the referee for comments which have helped us clarify this paper.
\section*{Data Availability}
The maser components analysed in this paper are presented in the text.
The raw interferometry data can be obtained from e-MERLIN on request.
The processed image cube can be provided on reasonable request to the corresponding author.

\begin{thebibliography}{}
\makeatletter
\relax
\def\mn@urlcharsother{\let\do\@makeother \do\$\do\&\do\#\do\^\do\_\do\%\do\~}
\def\mn@doi{\begingroup\mn@urlcharsother \@ifnextchar [ {\mn@doi@}
  {\mn@doi@[]}}
\def\mn@doi@[#1]#2{\def\@tempa{#1}\ifx\@tempa\@empty \href
  {http://dx.doi.org/#2} {doi:#2}\else \href {http://dx.doi.org/#2} {#1}\fi
  \endgroup}
\def\mn@eprint#1#2{\mn@eprint@#1:#2::\@nil}
\def\mn@eprint@arXiv#1{\href {http://arxiv.org/abs/#1} {{\tt arXiv:#1}}}
\def\mn@eprint@dblp#1{\href {http://dblp.uni-trier.de/rec/bibtex/#1.xml}
  {dblp:#1}}
\def\mn@eprint@#1:#2:#3:#4\@nil{\def\@tempa {#1}\def\@tempb {#2}\def\@tempc
  {#3}\ifx \@tempc \@empty \let \@tempc \@tempb \let \@tempb \@tempa \fi \ifx
  \@tempb \@empty \def\@tempb {arXiv}\fi \@ifundefined
  {mn@eprint@\@tempb}{\@tempb:\@tempc}{\expandafter \expandafter \csname
  mn@eprint@\@tempb\endcsname \expandafter{\@tempc}}}

\bibitem[\protect\citeauthoryear{{Arce}, {Shepherd}, {Gueth}, {Lee},
  {Bachiller}, {Rosen}  \& {Beuther}}{{Arce} et~al.}{2007}]{Arce2007}
{Arce} H.~G.,  {Shepherd} D.,  {Gueth} F.,  {Lee} C.~F.,  {Bachiller} R.,
  {Rosen} A.,   {Beuther} H.,  2007, in {Reipurth} B.,  {Jewitt} D.,   {Keil}
  K.,  eds, Protostars and Planets V. p.~245 (\mn@eprint {arXiv}
  {astro-ph/0603071})

\bibitem[\protect\citeauthoryear{Baars, Genzel, Pauliny-Toth  \& Witzel}{Baars
  et~al.}{1977}]{Baars1977}
Baars J. W.~M.,  Genzel R.,  Pauliny-Toth I. I.~K.,   Witzel A.,  1977, A\&A,
  61, 99

\bibitem[\protect\citeauthoryear{{Bartkiewicz}, {Szymczak}, {van Langevelde},
  {Richards}  \& {Pihlstr{\"o}m}}{{Bartkiewicz} et~al.}{2009}]{Bartkiewicz2009}
{Bartkiewicz} A.,  {Szymczak} M.,  {van Langevelde} H.~J.,  {Richards}
  A.~M.~S.,   {Pihlstr{\"o}m} Y.~M.,  2009, \mn@doi [A\&A]
  {10.1051/0004-6361/200912250}, \href
  {https://ui.adsabs.harvard.edu/abs/2009A&A...502..155B} {502, 155}

\bibitem[\protect\citeauthoryear{{Bartkiewicz}, {Szymczak}  \& {van
  Langevelde}}{{Bartkiewicz} et~al.}{2016}]{Bartkiewicz2016}
{Bartkiewicz} A.,  {Szymczak} M.,   {van Langevelde} H.~J.,  2016, \mn@doi
  [\aap] {10.1051/0004-6361/201527541}, \href
  {https://ui.adsabs.harvard.edu/abs/2016A&A...587A.104B} {587, A104}

\bibitem[\protect\citeauthoryear{Beichman \& Neugebauer}{Beichman \&
  Neugebauer}{1988}]{IRAS1988}
Beichman C.~A.,  Neugebauer G.,  eds, 1988, {Explanatory Supplement.}  Infrared
  astronomical satellite (IRAS) catalogs and atlases. Vol. 1.
IRAS Project Office

\bibitem[\protect\citeauthoryear{{Beuther} \& {Schilke}}{{Beuther} \&
  {Schilke}}{2004}]{Beuther2004}
{Beuther} H.,  {Schilke} P.,  2004, \mn@doi [Science]
  {10.1126/science.1094014}, \href
  {https://ui.adsabs.harvard.edu/abs/2004Sci...303.1167B} {303, 1167}

\bibitem[\protect\citeauthoryear{{Bloemhof}}{{Bloemhof}}{2000}]{Bloemhof20}
{Bloemhof} E.~E.,  2000, \mn@doi [\apj] {10.1086/308714}, \href
  {https://ui.adsabs.harvard.edu/abs/2000ApJ...533..893B} {533, 893}

\bibitem[\protect\citeauthoryear{{Bonnell}, {Bate}  \& {Zinnecker}}{{Bonnell}
  et~al.}{1998}]{Bonnell1998}
{Bonnell} I.~A.,  {Bate} M.~R.,   {Zinnecker} H.,  1998, \mn@doi [MNRAS]
  {10.1046/j.1365-8711.1998.01590.x}, \href
  {https://ui.adsabs.harvard.edu/abs/1998MNRAS.298...93B} {298, 93}

\bibitem[\protect\citeauthoryear{{Breen}, {Ellingsen}, {Caswell}  \&
  {Lewis}}{{Breen} et~al.}{2010}]{Breen2010}
{Breen} S.~L.,  {Ellingsen} S.~P.,  {Caswell} J.~L.,   {Lewis} B.~E.,  2010,
  \mn@doi [MNRAS] {10.1111/j.1365-2966.2009.15831.x}, \href
  {https://ui.adsabs.harvard.edu/abs/2010MNRAS.401.2219B} {401, 2219}

\bibitem[\protect\citeauthoryear{{Breen}, {Ellingsen}, {Contreras}, {Green},
  {Caswell}, {Stevens}, {Dawson}  \& {Voronkov}}{{Breen}
  et~al.}{2013}]{Breen2013}
{Breen} S.~L.,  {Ellingsen} S.~P.,  {Contreras} Y.,  {Green} J.~A.,  {Caswell}
  J.~L.,  {Stevens} J.~B.,  {Dawson} J.~R.,   {Voronkov} M.~A.,  2013, \mn@doi
  [MNRAS] {10.1093/mnras/stt1315}, \href
  {https://ui.adsabs.harvard.edu/abs/2013MNRAS.435..524B} {435, 524}

\bibitem[\protect\citeauthoryear{{Caswell}}{{Caswell}}{2009}]{Caswell2009}
{Caswell} J.~L.,  2009, \mn@doi [PASA] {10.1071/AS09013}, \href
  {https://ui.adsabs.harvard.edu/abs/2009PASA...26..454C} {26, 454}

\bibitem[\protect\citeauthoryear{{Cragg}, {Johns}, {Godfrey}  \&
  {Brown}}{{Cragg} et~al.}{1992}]{Cragg1992}
{Cragg} D.~M.,  {Johns} K.~P.,  {Godfrey} P.~D.,   {Brown} R.~D.,  1992,
  \mn@doi [MNRAS] {10.1093/mnras/259.1.203}, \href
  {https://ui.adsabs.harvard.edu/abs/1992MNRAS.259..203C} {259, 203}

\bibitem[\protect\citeauthoryear{{Darwish}, {Edris}, {Richards}, {Etoka},
  {Saad}, {Beheary}  \& {Fuller}}{{Darwish} et~al.}{2020}]{Darwish2020}
{Darwish} M.~S.,  {Edris} K.~A.,  {Richards} A.~M.~S.,  {Etoka} S.,  {Saad}
  M.~S.,  {Beheary} M.~M.,   {Fuller} G.~A.,  2020, \mn@doi [MNRAS]
  {10.1093/mnras/staa574}, \href
  {https://ui.adsabs.harvard.edu/abs/2020MNRAS.493.4442D} {493, 4442}

\bibitem[\protect\citeauthoryear{{De Villiers} et~al.,}{{De Villiers}
  et~al.}{2015}]{deVilliers2015}
{De Villiers} H.~M.,  et~al., 2015, \mn@doi [MNRAS] {10.1093/mnras/stv173},
  \href {https://ui.adsabs.harvard.edu/abs/2015MNRAS.449..119D} {449, 119}

\bibitem[\protect\citeauthoryear{{{Diamond, P. J. et al.}}}{{{Diamond, P. J. et
  al.}}}{2003}]{MUG}
{{Diamond, P. J. et al.}} 2003, {MERLIN User Guide}.
\url {http://www.merlin.ac.uk/user_guide/}

\bibitem[\protect\citeauthoryear{{Edris}, {Fuller}, {Cohen}  \&
  {Etoka}}{{Edris} et~al.}{2005}]{Edris2005}
{Edris} K.~A.,  {Fuller} G.~A.,  {Cohen} R.~J.,   {Etoka} S.,  2005, \mn@doi
  [A\&A] {10.1051/0004-6361:20041872}, \href
  {https://ui.adsabs.harvard.edu/abs/2005A&A...434..213E} {434, 213}

\bibitem[\protect\citeauthoryear{{Edris}, {Fuller}  \& {Cohen}}{{Edris}
  et~al.}{2007}]{Edris2007}
{Edris} K.~A.,  {Fuller} G.~A.,   {Cohen} R.~J.,  2007, \mn@doi [A\&A]
  {10.1051/0004-6361:20066280}, \href
  {https://ui.adsabs.harvard.edu/abs/2007A&A...465..865E} {465, 865}

\bibitem[\protect\citeauthoryear{{Fitzgibbon}, {Pilu}  \&
  {Fisher}}{{Fitzgibbon} et~al.}{1999}]{f99}
{Fitzgibbon} A.,  {Pilu} M.,   {Fisher} R.~B.,  1999, IEEE Trans. Pattern Anal.
  Mach. Intell., 21, 476

\bibitem[\protect\citeauthoryear{{G{\'o}mez-Ruiz}, {Kurtz}, {Araya}, {Hofner}
  \& {Loinard}}{{G{\'o}mez-Ruiz} et~al.}{2016}]{GomezRuiz2016}
{G{\'o}mez-Ruiz} A.~I.,  {Kurtz} S.~E.,  {Araya} E.~D.,  {Hofner} P.,
  {Loinard} L.,  2016, \mn@doi [ApJs] {10.3847/0067-0049/222/2/18}, \href
  {https://ui.adsabs.harvard.edu/abs/2016ApJS..222...18G} {222, 18}

\bibitem[\protect\citeauthoryear{{Gray}}{{Gray}}{2012}]{Gray2012}
{Gray} M.,  2012, {Maser Sources in Astrophysics}.
Cambridge University Press, UK

\bibitem[\protect\citeauthoryear{{Keto} \& {Zhang}}{{Keto} \&
  {Zhang}}{2010}]{Keto2010}
{Keto} E.,  {Zhang} Q.,  2010, \mn@doi [MNRAS]
  {10.1111/j.1365-2966.2010.16672.x}, \href
  {https://ui.adsabs.harvard.edu/abs/2010MNRAS.406..102K} {406, 102}

\bibitem[\protect\citeauthoryear{{Kong}, {Arce}, {Maureira}, {Caselli}, {Tan}
  \& {Fontani}}{{Kong} et~al.}{2019}]{Kong2019}
{Kong} S.,  {Arce} H.~G.,  {Maureira} M.~J.,  {Caselli} P.,  {Tan} J.~C.,
  {Fontani} F.,  2019, \mn@doi [ApJ] {10.3847/1538-4357/ab07b9}, \href
  {https://ui.adsabs.harvard.edu/abs/2019ApJ...874..104K} {874, 104}

\bibitem[\protect\citeauthoryear{{Kong}, {Arce}, {Shirley}  \&
  {Glasgow}}{{Kong} et~al.}{2021}]{Kong2021}
{Kong} S.,  {Arce} H.~G.,  {Shirley} Y.,   {Glasgow} C.,  2021, \mn@doi [\apj]
  {10.3847/1538-4357/abefe7}, \href
  {https://ui.adsabs.harvard.edu/abs/2021ApJ...912..156K} {912, 156}

\bibitem[\protect\citeauthoryear{{Kurtz}}{{Kurtz}}{2005}]{Kurtz2005}
{Kurtz} S.,  2005, in {Lis} D.~C.,  {Blake} G.~A.,   {Herbst} E.,  eds,
  Proceedings of the International Astronomical Union Vol. 231, Astrochemistry:
  Recent Successes and Current Challenges. pp 47--56,
  \mn@doi{10.1017/S1743921306007034}

\bibitem[\protect\citeauthoryear{{McKee} \& {Tan}}{{McKee} \&
  {Tan}}{2003}]{McKee2003}
{McKee} C.~F.,  {Tan} J.~C.,  2003, \mn@doi [ApJ] {10.1086/346149}, \href
  {https://ui.adsabs.harvard.edu/abs/2003ApJ...585..850M} {585, 850}

\bibitem[\protect\citeauthoryear{{Molinari}, {Brand}, {Cesaroni}  \&
  {Palla}}{{Molinari} et~al.}{1996}]{Molinari1996}
{Molinari} S.,  {Brand} J.,  {Cesaroni} R.,   {Palla} F.,  1996, A\&A, \href
  {https://ui.adsabs.harvard.edu/abs/1996A&A...308..573M} {308, 573}

\bibitem[\protect\citeauthoryear{{Molinari}, {Brand}, {Cesaroni}, {Palla}  \&
  {Palumbo}}{{Molinari} et~al.}{1998}]{Molinari1998}
{Molinari} S.,  {Brand} J.,  {Cesaroni} R.,  {Palla} F.,   {Palumbo} G.~G.~C.,
  1998, A\&A, \href {https://ui.adsabs.harvard.edu/abs/1998A&A...336..339M}
  {336, 339}

\bibitem[\protect\citeauthoryear{{Moscadelli}, {Cesaroni}  \&
  {Rioja}}{{Moscadelli} et~al.}{2000}]{Moscadelli2000}
{Moscadelli} L.,  {Cesaroni} R.,   {Rioja} M.~J.,  2000, A\&A, \href
  {https://ui.adsabs.harvard.edu/abs/2000A&A...360..663M} {360, 663}

\bibitem[\protect\citeauthoryear{{Moscadelli}, {Cesaroni}  \&
  {Rioja}}{{Moscadelli} et~al.}{2005}]{Moscadelli2005}
{Moscadelli} L.,  {Cesaroni} R.,   {Rioja} M.~J.,  2005, \mn@doi [\aap]
  {10.1051/0004-6361:20052685}, \href
  {https://ui.adsabs.harvard.edu/abs/2005A&A...438..889M} {438, 889}

\bibitem[\protect\citeauthoryear{{Norris} et~al.,}{{Norris}
  et~al.}{1998}]{Norris1998}
{Norris} R.~P.,  et~al., 1998, \mn@doi [ApJ] {10.1086/306373}, \href
  {https://ui.adsabs.harvard.edu/abs/1998ApJ...508..275N} {508, 275}

\bibitem[\protect\citeauthoryear{{Palla}, {Brand}, {Cesaroni}, {Comoretto}  \&
  {Felli}}{{Palla} et~al.}{1991}]{Palla1991}
{Palla} F.,  {Brand} J.,  {Cesaroni} R.,  {Comoretto} G.,   {Felli} M.,  1991,
  A\&A, \href {https://ui.adsabs.harvard.edu/abs/1991A&A...246..249P} {246,
  249}

\bibitem[\protect\citeauthoryear{{Reid} \& {Wilson}}{{Reid} \&
  {Wilson}}{2006}]{Reid2006}
{Reid} M.~A.,  {Wilson} C.~D.,  2006, \mn@doi [\apj] {10.1086/507019}, \href
  {https://ui.adsabs.harvard.edu/abs/2006ApJ...650..970R} {650, 970}

\bibitem[\protect\citeauthoryear{{Reid} et~al.,}{{Reid}
  et~al.}{2019}]{Reid2019}
{Reid} M.~J.,  et~al., 2019, \mn@doi [\apj] {10.3847/1538-4357/ab4a11}, \href
  {https://ui.adsabs.harvard.edu/abs/2019ApJ...885..131R} {885, 131}

\bibitem[\protect\citeauthoryear{{Rivilla}, {Mart{\'\i}n-Pintado},
  {Jim{\'e}nez-Serra}  \& {Rodr{\'\i}guez-Franco}}{{Rivilla}
  et~al.}{2013}]{Rivilla2013}
{Rivilla} V.~M.,  {Mart{\'\i}n-Pintado} J.,  {Jim{\'e}nez-Serra} I.,
  {Rodr{\'\i}guez-Franco} A.,  2013, \mn@doi [\aap]
  {10.1051/0004-6361/201117487}, \href
  {https://ui.adsabs.harvard.edu/abs/2013A&A...554A..48R} {554, A48}

\bibitem[\protect\citeauthoryear{{Seifried}, {Pudritz}, {Banerjee}, {Duffin}
  \& {Klessen}}{{Seifried} et~al.}{2012}]{Seifried2012}
{Seifried} D.,  {Pudritz} R.~E.,  {Banerjee} R.,  {Duffin} D.,   {Klessen}
  R.~S.,  2012, \mn@doi [\mnras] {10.1111/j.1365-2966.2012.20610.x}, \href
  {https://ui.adsabs.harvard.edu/abs/2012MNRAS.422..347S} {422, 347}

\bibitem[\protect\citeauthoryear{{Seifried}, {S{\'a}nchez-Monge}, {Walch}  \&
  {Banerjee}}{{Seifried} et~al.}{2016}]{Seifried2016}
{Seifried} D.,  {S{\'a}nchez-Monge} {\'A}.,  {Walch} S.,   {Banerjee} R.,
  2016, \mn@doi [MNRAS] {10.1093/mnras/stw785}, \href
  {https://ui.adsabs.harvard.edu/abs/2016MNRAS.459.1892S} {459, 1892}

\bibitem[\protect\citeauthoryear{Thomasson}{Thomasson}{1986}]{Thomasson1986}
Thomasson P.,  1986, Quat. Jl. Royal Astron. Soc., 27, 413

\bibitem[\protect\citeauthoryear{{Torstensson}, {van Langevelde}, {Vlemmings}
  \& {Bourke}}{{Torstensson} et~al.}{2011}]{Torstensson2011}
{Torstensson} K.~J.~E.,  {van Langevelde} H.~J.,  {Vlemmings} W.~H.~T.,
  {Bourke} S.,  2011, \mn@doi [\aap] {10.1051/0004-6361/201015583}, \href
  {https://ui.adsabs.harvard.edu/abs/2011A&A...526A..38T} {526, A38}

\bibitem[\protect\citeauthoryear{{Varricatt}, {Davis}, {Ramsay}  \&
  {Todd}}{{Varricatt} et~al.}{2010}]{Varricatt2010}
{Varricatt} W.~P.,  {Davis} C.~J.,  {Ramsay} S.,   {Todd} S.~P.,  2010, \mn@doi
  [MNRAS] {10.1111/j.1365-2966.2010.16356.x}, \href
  {https://ui.adsabs.harvard.edu/abs/2010MNRAS.404..661V} {404, 661}

\bibitem[\protect\citeauthoryear{{Varricatt}, {Wouterloot}, {Ramsay}  \&
  {Davis}}{{Varricatt} et~al.}{2018}]{Varricatt2018}
{Varricatt} W.~P.,  {Wouterloot} J.~G.~A.,  {Ramsay} S.~K.,   {Davis} C.~J.,
  2018, \mn@doi [MNRAS] {10.1093/mnras/sty2099}, \href
  {https://ui.adsabs.harvard.edu/abs/2018MNRAS.480.4231V} {480, 4231}

\bibitem[\protect\citeauthoryear{{Zhang}, {Hunter}, {Brand}, {Sridharan},
  {Cesaroni}, {Molinari}, {Wang}  \& {Kramer}}{{Zhang}
  et~al.}{2005}]{Zhang2005}
{Zhang} Q.,  {Hunter} T.~R.,  {Brand} J.,  {Sridharan} T.~K.,  {Cesaroni} R.,
  {Molinari} S.,  {Wang} J.,   {Kramer} M.,  2005, \mn@doi [ApJ]
  {10.1086/429660}, \href
  {https://ui.adsabs.harvard.edu/abs/2005ApJ...625..864Z} {625, 864}

\makeatother
\end{thebibliography}

\newpage
\appendix
\onecolumn
\section{Further consideration of Keplerian models}
\label{sec_Kepler4}

\begin{figure*}
\includegraphics[width=15cm]{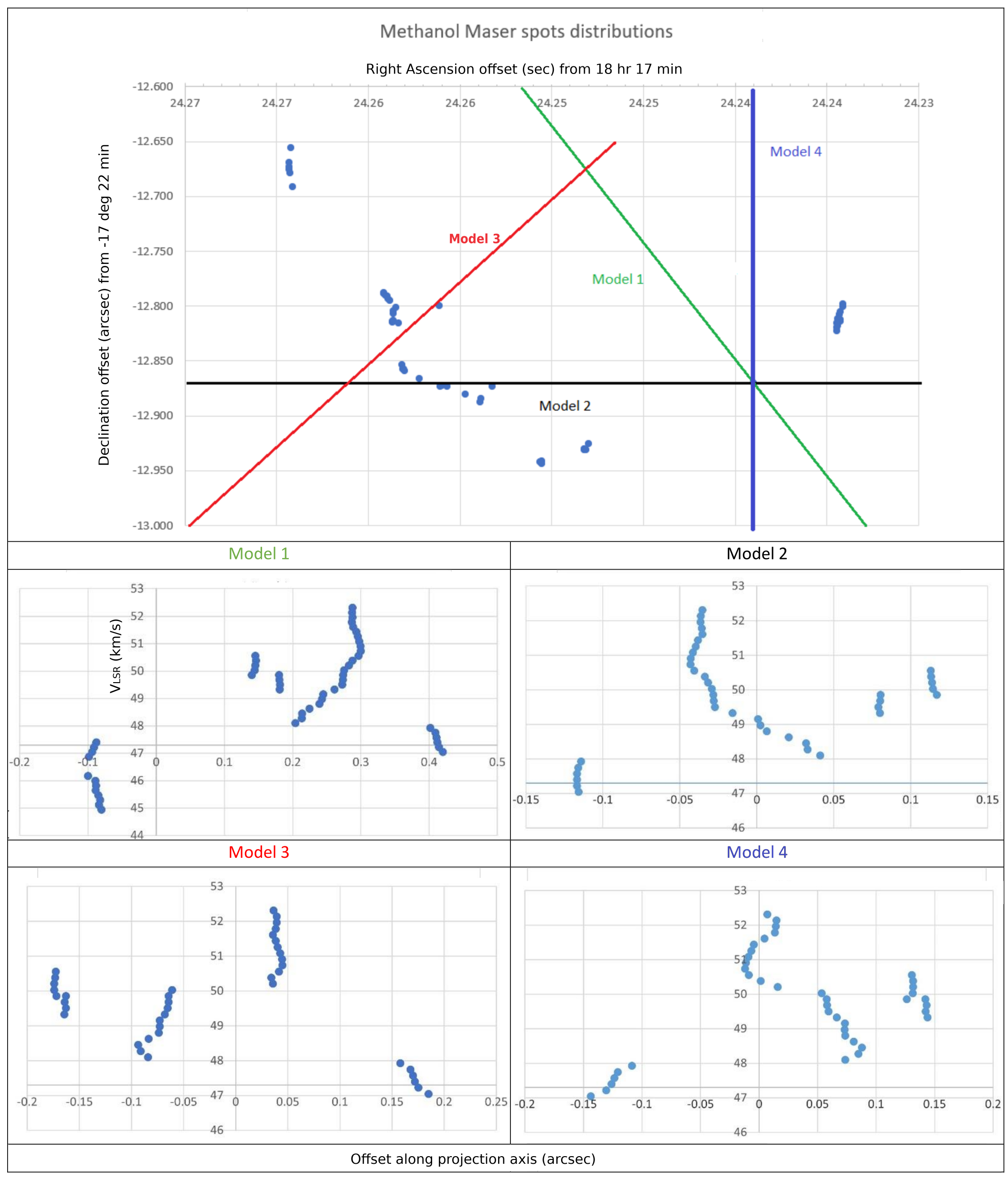}
\caption{Top panel: The distribution of 6.7 GHz methanol masers as in Fig.~\ref{fig-2}. The numbered lines correspond to the axes used to project the masers in the position-velocity diagrams in the lower panels. Measuring the axes angles from North through East, Models 1, 2, 3 and 4 are at 35$^{\circ}$, 90$^{\circ}$ (the R.A. axis), --35$^{\circ}$ and 0$^{\circ}$ (the Dec. axis). The bottom panels show the relative positions and velocities of the spots projected onto these axes. It can be seen that there is no orientation which would allow the maser spots to follow a Keplerian curve as in Fig.~\ref{fig-6}, for any enclosed mass or origin.}
\label{fig-A4}
\end{figure*}

\bsp	
\label{lastpage}
\end{document}